\PassOptionsToPackage{unicode}{hyperref}
\PassOptionsToPackage{hyphens}{url}
\PassOptionsToPackage{dvipsnames,svgnames,x11names}{xcolor}
\documentclass[
]{hdsr}

\usepackage{amsmath,amssymb}
\usepackage{iftex}
\ifPDFTeX
  \usepackage[T1]{fontenc}
  \usepackage[utf8]{inputenc}
  \usepackage{textcomp} % provide euro and other symbols
\else % if luatex or xetex
  \usepackage{unicode-math}
  \defaultfontfeatures{Scale=MatchLowercase}
  \defaultfontfeatures[\rmfamily]{Ligatures=TeX,Scale=1}
\fi
\usepackage{lmodern}
\ifPDFTeX\else  
\fi
\IfFileExists{upquote.sty}{\usepackage{upquote}}{}
\IfFileExists{microtype.sty}{% use microtype if available
  \usepackage[]{microtype}
  \UseMicrotypeSet[protrusion]{basicmath} % disable protrusion for tt fonts
}{}
\makeatletter
\@ifundefined{KOMAClassName}{% if non-KOMA class
  \IfFileExists{parskip.sty}{%
    \usepackage{parskip}
  }{% else
    \setlength{\parindent}{0pt}
    \setlength{\parskip}{6pt plus 2pt minus 1pt}}
}{% if KOMA class
  \KOMAoptions{parskip=half}}
\makeatother
\usepackage{xcolor}
\ifx\paragraph\undefined\else
  \let\oldparagraph\paragraph
  \renewcommand{\paragraph}[1]{\oldparagraph{#1}\mbox{}}
\fi
\ifx\subparagraph\undefined\else
  \let\oldsubparagraph\subparagraph
  \renewcommand{\subparagraph}[1]{\oldsubparagraph{#1}\mbox{}}
\fi

\usepackage{color}
\usepackage{fancyvrb}

\DefineVerbatimEnvironment{Highlighting}{Verbatim}{commandchars=\\\{\}}
\usepackage{framed}
\definecolor{shadecolor}{RGB}{241,243,245}
\newenvironment{Shaded}{\begin{snugshade}}{\end{snugshade}}

\newcommand{\AttributeTok}[1]{\textcolor[rgb]{0.40,0.45,0.13}{#1}}

\newcommand{\BuiltInTok}[1]{\textcolor[rgb]{0.00,0.23,0.31}{#1}}

\newcommand{\CommentTok}[1]{\textcolor[rgb]{0.37,0.37,0.37}{#1}}

\newcommand{\ControlFlowTok}[1]{\textcolor[rgb]{0.00,0.23,0.31}{#1}}

\newcommand{\DecValTok}[1]{\textcolor[rgb]{0.68,0.00,0.00}{#1}}

\newcommand{\FloatTok}[1]{\textcolor[rgb]{0.68,0.00,0.00}{#1}}
\newcommand{\FunctionTok}[1]{\textcolor[rgb]{0.28,0.35,0.67}{#1}}
\newcommand{\ImportTok}[1]{\textcolor[rgb]{0.00,0.46,0.62}{#1}}

\newcommand{\KeywordTok}[1]{\textcolor[rgb]{0.00,0.23,0.31}{#1}}
\newcommand{\NormalTok}[1]{\textcolor[rgb]{0.00,0.23,0.31}{#1}}
\newcommand{\OperatorTok}[1]{\textcolor[rgb]{0.37,0.37,0.37}{#1}}
\newcommand{\OtherTok}[1]{\textcolor[rgb]{0.00,0.23,0.31}{#1}}

\newcommand{\SpecialCharTok}[1]{\textcolor[rgb]{0.37,0.37,0.37}{#1}}

\newcommand{\StringTok}[1]{\textcolor[rgb]{0.13,0.47,0.30}{#1}}
\newcommand{\VariableTok}[1]{\textcolor[rgb]{0.07,0.07,0.07}{#1}}

\providecommand{\tightlist}{%
  \setlength{\itemsep}{0pt}\setlength{\parskip}{0pt}}\usepackage{longtable,booktabs,array}
\usepackage{calc} % for calculating minipage widths
\usepackage{etoolbox}
\makeatletter
\patchcmd\longtable{\par}{\if@noskipsec\mbox{}\fi\par}{}{}
\makeatother
\IfFileExists{footnotehyper.sty}{\usepackage{footnotehyper}}{\usepackage{footnote}}
\makesavenoteenv{longtable}
\usepackage{graphicx}
\makeatletter
\def\maxwidth{\ifdim\Gin@nat@width>\linewidth\linewidth\else\Gin@nat@width\fi}
\def\maxheight{\ifdim\Gin@nat@height>\textheight\textheight\else\Gin@nat@height\fi}
\makeatother
\setkeys{Gin}{width=\maxwidth,height=\maxheight,keepaspectratio}
\makeatletter
\def\fps@figure{htbp}
\makeatother
\newlength{\cslhangindent}
\newlength{\csllabelwidth}
\newlength{\cslentryspacingunit} % times entry-spacing
\newenvironment{CSLReferences}[2] % #1 hanging-ident, #2 entry spacing
 {% don't indent paragraphs
  \setlength{\parindent}{0pt}
  \ifodd #1
  \let\oldpar\par
  \def\par{\hangindent=\cslhangindent\oldpar}
  \fi
  \setlength{\parskip}{#2\cslentryspacingunit}
 }%
 {}
\usepackage{calc}

\usepackage{booktabs}
\usepackage{longtable}
\usepackage{array}
\usepackage{multirow}
\usepackage{wrapfig}
\usepackage{float}
\usepackage{colortbl}
\usepackage{pdflscape}
\usepackage{tabu}
\usepackage{threeparttable}
\usepackage{threeparttablex}
\usepackage[normalem]{ulem}
\usepackage{makecell}
\usepackage{xcolor}
\graphicspath{{figs/}}
\makeatletter
\@ifpackageloaded{caption}{}{\usepackage{caption}}
\AtBeginDocument{%
\ifdefined\contentsname
  \renewcommand*\contentsname{Table of contents}
\else
  \newcommand\contentsname{Table of contents}
\fi
\ifdefined\listfigurename
  \renewcommand*\listfigurename{List of Figures}
\else
  \newcommand\listfigurename{List of Figures}
\fi
\ifdefined\listtablename
  \renewcommand*\listtablename{List of Tables}
\else
  \newcommand\listtablename{List of Tables}
\fi
\ifdefined\figurename
  \renewcommand*\figurename{Figure}
\else
  \newcommand\figurename{Figure}
\fi
\ifdefined\tablename
  \renewcommand*\tablename{Table}
\else
  \newcommand\tablename{Table}
\fi
}
\@ifpackageloaded{float}{}{\usepackage{float}}
\floatstyle{ruled}
\@ifundefined{c@chapter}{\newfloat{codelisting}{h}{lop}}{\newfloat{codelisting}{h}{lop}[chapter]}
\floatname{codelisting}{Listing}

\makeatother
\makeatletter
\@ifpackageloaded{caption}{}{\usepackage{caption}}
\@ifpackageloaded{subcaption}{}{\usepackage{subcaption}}
\makeatother
\makeatletter
\@ifpackageloaded{tcolorbox}{}{\usepackage[skins,breakable]{tcolorbox}}
\makeatother
\makeatletter
\@ifundefined{shadecolor}{\definecolor{shadecolor}{rgb}{.97, .97, .97}}
\makeatother
\makeatletter
\@ifpackageloaded{tikz}{}{\usepackage{tikz}}
\makeatother
        \newcommand*\circled[1]{\tikz[baseline=(char.base)]{
          \node[shape=circle,draw,inner sep=1pt] (char) {{\scriptsize#1}};}}  
                  
\ifLuaTeX
  \usepackage{selnolig}  % disable illegal ligatures
\fi
\IfFileExists{bookmark.sty}{\usepackage{bookmark}}{\usepackage{hyperref}}
\IfFileExists{xurl.sty}{\usepackage{xurl}}{} % add URL line breaks if available
\hypersetup{
  pdftitle={Fostering better coding practices for data scientists},
  pdfauthor={Randall Pruim; Maria-Cristiana Gîrjău; Nicholas J. Horton},
  pdfkeywords={data acumen, data science, data science practice, data
science education, code quality, code style},
  colorlinks=true,
  linkcolor={blue},
  filecolor={Maroon},
  citecolor={Blue},
  urlcolor={Blue},
  pdfcreator={LaTeX via pandoc}}

\title{Fostering better coding practices for data scientists}
\begin{document}

% \graphicspath{{figs/}}

\newgeometry{bottom=1.5in}

%\volumeheader%
%{5}%
%{3}%
%{10.1162/99608f92.97c9f60f}

\begin{center}
\maketitle

% Start page numbering on second page. Must appear *after* \maketitle
\thispagestyle{empty}

 % % Authors and Affiliations
 %  \begin{tabular}{cc}
 %    First Author\upstairs{\affilone,*}, Second Author\upstairs{\affilone}, Third Author\upstairs{\affilthree}
 %   \\[0.25ex]
 %   {\small \upstairs{\affilone} Affiliation One} \\
 %   {\small \upstairs{\affiltwo} Affiliation Two} \\
 %   {\small \upstairs{\affilthree} Affiliation Three} \\
 %  \end{tabular}
 
\begin{tabular}{cc}
Randall Pruim\upstairs{\affilone}, Maria-Cristiana Gîrjău\upstairs{\affiltwo, \affilthree}, Nicholas J. Horton\upstairs{\affiltwo,*} \\[.25ex]
 {\small \upstairs{\affilone} Calvin University} \\
 {\small \upstairs{\affiltwo} Amherst College} \\
 {\small \upstairs{\affilthree} Columbia University} \\
\end{tabular} 

 % Replace with corresponding author email address
 \emails{
   \upstairs{*}nhorton@amherst.edu
   }
 \vspace*{0.4in}

\begin{abstract}
Many data science students and practitioners don't see the value in
making time to learn and adopt good coding practices as long as the code
``works''. However, code standards are an important part of modern data
science practice, and they play an essential role in the development of
data acumen. Good coding practices lead to more reliable code and save
more time than they cost, making them important even for beginners. We
believe that principled coding is vital for quality data science
practice. To effectively instill these practices within academic
programs, instructors and programs need to begin establishing these
practices early, to reinforce them often, and to hold themselves to a
higher standard while guiding students. We describe key aspects of good
coding practices for data science, illustrating with examples in R and
in Python, though similar standards are applicable to other software
environments. Practical coding guidelines are organized into a top ten
list.
\end{abstract}

\end{center}

\vspace*{0.15in}

\hspace{10pt}

\copyrightnotice

% Media summary should come next as an unnumbered section\ifdefined\Shaded\renewenvironment{Shaded}{\begin{tcolorbox}[frame hidden, enhanced, boxrule=0pt, breakable, interior hidden, sharp corners, borderline west={3pt}{0pt}{shadecolor}]}{\end{tcolorbox}}\fi

\hypertarget{media-summary}{%
\section*{Media summary}\label{media-summary}}
\addcontentsline{toc}{section}{Media summary}

Many data science students and practitioners are reluctant to adopt good
coding practices as long as the code ``works''. Yet meticulous attention
to detail is an important characteristic of a data scientist.

Code standards are an important part of modern data science, and they
play an essential role in ensuring the quality of data science in
research and in the workforce. Responsible coding practices lead to more
reliable code and save more time than they cost, making them important
even for beginners. We believe that principled coding is vital for
quality data science practice.

To effectively instill these habits of mind within academic programs,
instructors and programs need to begin establishing these practices
early, to reinforce them often, and to hold themselves to a higher
standard while guiding students. We describe key aspects of good coding
for data science, illustrating them with examples and motivation.
Practical coding guidelines are organized into a top ten list.

\hypertarget{introduction}{%
\section{Introduction}\label{introduction}}

Coding is an increasingly important part of statistical analyses
(Deboran Nolan and Temple Lang 2010; Hardin et al. 2021). The goal of
code is not just to solve an immediate analysis problem but also to
establish reusable workflows and to communicate. As projects and
analyses become more sophisticated, it's important that structures and
expectations be set to facilitate data science as a \emph{team sport} in
a sustainable and reproducible way (Horton et al. 2022).

Consensus reports (National Academies of Science, Engineering, and
Medicine 2018) have highlighted the importance of workflow and
reproducibility as a component of data science practice:

\begin{quote}
``Modern data science has at its core the creation of
workflows---pipelines of processes that combine simpler tools to solve
larger tasks. Documenting, incrementally improving, sharing, and
generalizing such workflows are an important part of data science
practice owing to the team nature of data science and broader
significance of scientific reproducibility and replicability.
Documenting and sharing workflows enable others to understand how data
have been used and refined and what steps were taken in an analysis
process. This can increase the confidence in results and improve trust
in the process as well as enable reuse of analyses or results in a
meaningful way (page 2-12).''
\end{quote}

Adhering to established coding standards is an important step in
fostering effective analyses and an important component of data acumen.
Unfortunately, attention to these issues has not been central to many
curricula in statistics and data science.

\hypertarget{why-bother}{%
\subsection{Why bother?}\label{why-bother}}

Many data science students, and some practitioners, do not make time to
learn and adopt good coding practices. As long as the code ``works'',
they are satisfied and ready to move on. But how do they know that ``it
works''? For data analysts, these issues are an important part of data
science practice, because it affects the bottom line:

\begin{center}
\textbf{Good coding practices lead to more reliable and maintainable code.}
\end{center}

It is easier to notice and fix errors in well written code. And it is
less likely that the errors occur in the first place if the authors are
using good practices. Trisovic et al. (2022) carried out a large-scale
study on code quality and found that 74\% of R files did not run
successfully. After incorporating some automated code cleaning targeting
``some of the most common execution errors'', 56\% still failed.

Even when the code is correct, following good coding practices makes the
code easier to read and understand, saving time and promoting good
communication among team members.

In a blog post, Lyman (2021) wrote:

\begin{quote}
The value of high-quality code can be difficult to communicate. Some
managers see it as a boondoggle, an expensive hobby for overly
fastidious programmers, since investing in code quality can slow
development over the short term and doesn't appear to alter the user
experience. But nothing could be further from the truth.
\end{quote}

Learning and consistently using good coding practices takes some effort
and some attention. But in the end, we agree that they are likely to
save far more time than they consume (Ball et al. 2022). Well written
code is more likely to be correct, saving the time of redoing things,
and easier to maintain, saving effort when it becomes necessary to
modify or adapt the code in the future. Collaborations with other team
members are likely to be more positive if the quality of individual
contributions is higher.

We are convinced that

\begin{center}
\textbf{Good coding practice is important, even for beginners.}
\end{center}

It is easier and more efficient to learn good coding practices as one
learns to program than to unlearn bad habits later. This makes it
especially important that the code beginners \emph{see} meets the
highest standards for coding practice. We can't expect beginners to
mimic these practices perfectly from the start, and we recommend
focusing student attention (and feedback) on just a few key coding
practices early on. But if they don't have a good model to emulate, we
are impeding their progress unnecessarily. As an additional benefit,
modeling good coding practices will make it easier for students (and
others) to learn not only good coding technique but also the concepts
and applications that the code is illustrating.

In this paper, we will motivate the importance of principled coding,
illustrate key aspects of good coding practices, and suggest ways that
these practices can be included in the data science and statistics
curriculum.

\hypertarget{prior-work}{%
\subsection{Prior work}\label{prior-work}}

We acknowledge that much of what we discuss is not novel, but it is
nonetheless important (and, we argue, under-appreciated and
under-emphasized).

Many calls for better coding practices and enumerations of such
practices exist. Computer science curricula have long emphasized these
practices beginning in introductory programming courses and continuing
throughout the curriculum (Keuning, Heeren, and Jeuring 2017; Borstler
et al. 2017), especially in courses like software engineering or in
capstone projects courses (e.g., Berkeley's CS169,
\url{https://bcourses.berkeley.edu/courses/1507976}). Stegeman et al.
(2014) and (2016) have described rubrics and assessment for code quality
in programming courses.

The importance of good coding practices is also recognized in industry
({``{Google Style Guide}''} 2019; Ghani 2022) and across the sciences
(Wilson et al. 2017; Aruliah et al. 2012; Filazzola and Lortie 2022) and
social sciences (Gentzkow and Shapiro 2022). Dogucu and Çetinkaya-Rundel
(2022) motivates the importance of code quality, style guides, file
organization, and related topics. Related work by Carey and Papin (2018)
that describes rules for new programmers has relevance for teaching data
analysis. Deborah Nolan and Stoudt (2021) offer a ``Dirty Dozen'' set of
helpful code recommendations, and Abouzekry (2012) provides ten tips for
better coding.

Code quality has been an area where some previous research has been
undertaken. Schulte (2008) introduced a block model to help study
comprehension of program components (atoms, blocks, relations, and
macrostructure). Keuning, Heeren, and Jeuring (2017) and Keuning,
Heeren, and Jeuring (2019) have explored other aspects of teaching code
quality.

While the particular coding practices enumerated vary some by author,
programming language, and application area, the overall message is
clear: Good coding practices are important across a wide range of
contexts to ensure that people, especially those working in teams, are
productive and that their work is reliable, maintainable, and
reproducible. Furthermore, there is broad agreement about the basic
contours of what constitutes good coding practice. Unfortunately, the
abundance of such calls indicates that practice continues to fall short
of principle.

\hypertarget{a-motivating-example}{%
\subsection{A Motivating Example}\label{a-motivating-example}}

An April 2021 twitter post (Meyer 2021) commented:

\begin{quote}
``It is really painful when taking a graduate level data science course
and the instructor's code is considerably below any acceptable standard
in the real world. Here is some real life code from a demo offered for
the current homework\ldots{}''
\end{quote}

\begin{Shaded}
\begin{Highlighting}[]
\NormalTok{    data }\OtherTok{\textless{}{-}} \FunctionTok{read.csv}\NormalTok{(fname)}
\NormalTok{    train }\OtherTok{\textless{}{-}}\NormalTok{ data[}\DecValTok{0}\SpecialCharTok{:}\NormalTok{(}\FunctionTok{length}\NormalTok{(data[,}\DecValTok{1}\NormalTok{])}\SpecialCharTok{{-}}\DecValTok{8}\NormalTok{),]}
\NormalTok{    train.ts }\OtherTok{\textless{}{-}} \FunctionTok{ts}\NormalTok{(train[,}\FunctionTok{c}\NormalTok{(}\DecValTok{2}\NormalTok{,}\DecValTok{3}\NormalTok{,}\DecValTok{4}\NormalTok{,}\DecValTok{5}\NormalTok{)],,}\AttributeTok{start=}\FunctionTok{c}\NormalTok{(}\DecValTok{2014}\NormalTok{,}\DecValTok{1}\NormalTok{),}\AttributeTok{freq=}\DecValTok{52}\NormalTok{)}
\NormalTok{    test }\OtherTok{\textless{}{-}}\NormalTok{ data[(}\FunctionTok{length}\NormalTok{(data[,}\DecValTok{1}\NormalTok{])}\SpecialCharTok{{-}}\DecValTok{7}\NormalTok{)}\SpecialCharTok{:}\FunctionTok{length}\NormalTok{(data[,}\DecValTok{1}\NormalTok{]),]}
\end{Highlighting}
\end{Shaded}

We concur that this code, while short, is exceptionally hard to read.
There are many ways that this could be improved, some of which are
demonstrated below.

\hypertarget{annotated-cell-2}{%
\label{annotated-cell-2}}%
\begin{Shaded}
\begin{Highlighting}[]
\CommentTok{\# R}
\NormalTok{n\_test }\OtherTok{\textless{}{-}} \DecValTok{8}
\NormalTok{data }\OtherTok{\textless{}{-}}\NormalTok{ readr}\SpecialCharTok{::}\FunctionTok{read\_csv}\NormalTok{(fname)  }\hspace*{\fill}\NormalTok{\circled{1}}
\NormalTok{train }\OtherTok{\textless{}{-}} \FunctionTok{head}\NormalTok{(data, }\SpecialCharTok{{-}}\NormalTok{ n\_test)}
\NormalTok{test }\OtherTok{\textless{}{-}} \FunctionTok{tail}\NormalTok{(data, n\_test)}
\NormalTok{train\_ts }\OtherTok{\textless{}{-}} \FunctionTok{ts}\NormalTok{(}
\NormalTok{  dplyr}\SpecialCharTok{::}\FunctionTok{select}\NormalTok{(train, }\DecValTok{2}\SpecialCharTok{:}\DecValTok{5}\NormalTok{), }\hspace*{\fill}\NormalTok{\circled{2}}
  \AttributeTok{start =} \FunctionTok{c}\NormalTok{(}\DecValTok{2014}\NormalTok{, }\DecValTok{1}\NormalTok{), }
  \AttributeTok{freq =} \DecValTok{52}\NormalTok{)}
\end{Highlighting}
\end{Shaded}

\begin{description}
\tightlist
\item[\circled{1}]
Alternatively we could load the entire \texttt{readr} package with
\texttt{library(readr)} and avoid the \texttt{::}. For demonstration
code, the explicit package reference can help the reader know which
functions come from which packages. When using packages that are already
familiar to the reader or when using many functions from the same
package, loading with \texttt{library()} is more appropriate. We will
demonstrate both styles in the various examples presented here.
\item[\circled{2}]
Ideally we would use column names rather than numerical indices here.
\texttt{dplyr::select()} would be more useful in that case, especially
in conjunction with functions like \texttt{matches()},
\texttt{contains()}, \texttt{begins\_with()}, etc.
\end{description}

The suggested revisions add white space to improve readability and
clarify the type of subsetting that is happening by taking advantage of
the \texttt{head()} and \texttt{tail()} functions. In addition to
improved formatting, we specify that the \texttt{select()} function is
coming from the \texttt{dplyr} package.\footnote{We could also have
  loaded the \texttt{dplyr} package with \texttt{library(dplyr)}, and we
  would have done so had we used several functions from that package.}
We note that there are still aspects of the code that are brittle (e.g.,
assumptions regarding the ordering of the five variables in the
dataset). The code snippet in the tweet does not provide enough context
to appropriately address these. Using the native pipe in R, we could
rearrange the nested tasks, making the order of operations more
transparent. We will see examples of this shortly, and of a similar
approach called method chaining (see Augspurger 2016), in Python.

The revised code is easier to read and understand (and maintain) than
the original.

We believe that examples of this kind are all too common. This
particular example is compelling because it shows that students notice
the quality of the code instructors present and motivates why academics
need to live up to industry standards.

\hypertarget{establishing-good-coding-practices}{%
\section{Establishing Good Coding
Practices}\label{establishing-good-coding-practices}}

\hypertarget{sec-fourc}{%
\subsection{The Four C's}\label{sec-fourc}}

As mentioned in the introduction, many lists of good coding practices
have been published. These lists can provide useful guidance as one
progresses --- or, in the case of an instructor, leads students in their
progression --- toward better coding practices.

In addition to providing students with specific coding guidelines like
our top ten list below, we think it is also important that students
understand the higher level goals that the specific guidelines are
intended to support. These higher level goals take into account that
computer code simultaneously communicates both to humans and to the
computer and provide a framework for establishing a set of specific
coding practices. We describe these as the four C's for good code:

\begin{itemize}
\item
  \textbf{Correctness:} It is important that the code be correct so that
  the computer does what is intended. This, in itself, is not profound.
  But we emphasize two things about correctness: First, correctness is a
  \emph{necessary} but \emph{insufficient} metric for good code. Second,
  the other goals support and promote correctness.
\item
  \textbf{Clarity:} It is also important that the code be clear, so that
  humans reading and writing the code can tell what it is intended to
  do, and easily make modifications as necessary. (This advice applies
  both to other humans and to the same human at some later date.)
\item
  \textbf{Containment:} It is helpful if the code is appropriately
  contained, to keep separate things that should be separate and
  together things that should be together. Vartanian (2022) refers to
  this idea as ``low coupling, high cohesion.'' Other authors refer to
  this as modularization. Proper containment includes things like
  preferring a data frame over several individual vectors, using
  functions to contain reusable code, and keeping code used across files
  or projects in a module or package.
\item
  \textbf{Consistency:} Finally, it is useful if code exhibits internal
  consistency of style, naming conventions, and other coding practices.
\end{itemize}

The specific guidelines outlined below serve as concrete advice for
developing practices that promote creating code that satisfies the 4
C's.

\hypertarget{a-progressive-approach}{%
\subsection{A Progressive Approach}\label{a-progressive-approach}}

Before revealing our top ten list, we want to emphasize the importance
of taking a progressive approach, both for oneself and for students.
Developing good coding practices takes time and attention. Any list of
coding guidelines can exceed the available cognitive resources to take
them on. As is true for any behavioral intervention, change takes time
and effort.

This advice about developing a growth mindset that proposes slow and
steady change (Dana Center 2021) applied to the authors as well. We are
often dismayed and chagrined at the quality of the code we provided
students five or ten years ago, and we are constantly updating our own
coding examples to improve them.

\hypertarget{top-ten-list}{%
\subsection{Top Ten List}\label{top-ten-list}}

While the four C's establish important goals for code, they do little to
specify how the goals might be achieved. The list of ten guidelines
below provide some additional specificity and are ordered roughly in the
progression we encourage our students to develop them. By the completion
of an undergraduate data science program, we would expect all students
to be comfortable practicing all of the items on this list, which we are
confident will also serve data science practitioners well throughout
their careers.

\hypertarget{sec-good-names}{%
\subsubsection{\texorpdfstring{\textbf{Choose good
names.}}{Choose good names.}}\label{sec-good-names}}

Wilson et al. (2017), Lyman (2021), and Jenny Bryan (2015) note the
importance of giving variables and functions meaningful names as a way
to clarify the code and make it easier to read. But naming things can be
difficult (Fowler 2009), especially early in a project (when the scope
may not yet be clear) and for novice coders. Having a set of general
purpose guidelines to narrow the choices and using these across multiple
projects can assist greatly in the selection of names.

\begin{enumerate}
\def\labelenumi{\alph{enumi}.}
\item
  The length of names should be proportional to their scope.

  The more distance (measured in terms of number of people, human time,
  and lines of code) between definition and use, the more important it
  is that a name communicates clearly. The use of a single-character
  variable name may be acceptable as an index variable of limited scope
  or as a placeholder argument for a simple function. But even in these
  cases, a name that reflects what the indexing or placeholder
  represents is often preferred. Abbreviations are a two-edged sword.
  Used consistently by a community or team that is familiar with their
  meanings, they can help reduce the length of names. They can also
  become inscrutable to those less familiar with the project. Keeping a
  digest of abbreviations used can be very helpful.
\item
  Use capitalization consistently.

  There are no absolute standards here, but adopting a strong local
  convention can help avoid errors. It is also wise to avoid using two
  names that differ only in their case. Such names are easily swapped
  for one another and are difficult to read aloud. Students should be
  introduced to naming schemes such as \texttt{camelCase},
  \texttt{PascalCase}, and \texttt{snake\_case}, but encouraged to stick
  with one of these as much as possible.\footnote{Unfortunately, R is
    not very helpful in this regard. Many naming inconsistencies exist
    in core R functions. The tidyverse has paid much more careful
    attention to naming conventions, and those who adopt the tidyverse
    tool kit should follow its naming conventions as much as possible.}
  We recommend avoiding the dot (.) as a delimiter in R since it serves
  another purpose in R's S3 generic system (and in other programming
  languages as well).
\item
  Avoid nondescript names.

  The ubiquitous \texttt{d} or \texttt{df} as the name for a data frame
  is a common example of a nondescript name. In a data analysis
  situation, the data and their provenance are important and the name of
  the variable containing the data being analyzed should communicate
  something about the data. Even in ``generic'' examples, descriptive
  names can be helpful.

\begin{Shaded}
\begin{Highlighting}[]
\CommentTok{\# original R}
\NormalTok{d }\OtherTok{\textless{}{-}} \FunctionTok{read.csv}\NormalTok{(}\StringTok{"study{-}2023.csv"}\NormalTok{)}
\NormalTok{d2 }\OtherTok{\textless{}{-}}\NormalTok{ d[ d}\SpecialCharTok{$}\NormalTok{age }\SpecialCharTok{\textgreater{}=} \DecValTok{18}\NormalTok{, ]}
\end{Highlighting}
\end{Shaded}

\hypertarget{annotated-cell-2}{%
\label{annotated-cell-2}}%
\begin{Shaded}
\begin{Highlighting}[]
\CommentTok{\# improved R}
\NormalTok{AllSubjects2023 }\OtherTok{\textless{}{-}}\NormalTok{ readr}\SpecialCharTok{::}\FunctionTok{read\_csv}\NormalTok{(}\StringTok{"study{-}2023.csv"}\NormalTok{)  }\hspace*{\fill}\NormalTok{\circled{1}}
\NormalTok{Adults2023 }\OtherTok{\textless{}{-}}\NormalTok{ AllSubjects2023 }\SpecialCharTok{|\textgreater{}} \FunctionTok{filter}\NormalTok{(age }\SpecialCharTok{\textgreater{}=} \DecValTok{18}\NormalTok{)    }\hspace*{\fill}\NormalTok{\circled{2}}
\end{Highlighting}
\end{Shaded}

  \begin{description}
  \tightlist
  \item[\circled{1}]
  For most files of modest size, the use of \texttt{readr::read\_csv()}
  rather than \texttt{read.csv()} will make very little difference
  although the format of the object returned may be slightly different.
  But \texttt{read\_csv()} is faster, more consistent across operating
  systems, and a bit more predictable.
  \item[\circled{2}]
  The pipe (\texttt{\textbar{}\textgreater{}}) passes its left hand side
  as the first argument of the function on the right hand side.
  \end{description}

  In situations where a variable is intended for frequent reassignment
  (in a loop, for example) or for names of formal arguments in
  functions, the considerations may be a little different. In these
  cases the name may say more about the role or hint at the intended use
  or data type.
\end{enumerate}

External resources, like files, benefit from consistent naming patterns
as well.

\begin{enumerate}
\def\labelenumi{\alph{enumi}.}
\item
  Use delimiters to make parsing easier for humans and computers.

  When one adopts mixed case or uses underscores or some other
  delimiter, consistent use of delimiters makes it easier for people to
  remember the name and opens the possibility of algorithmic processing
  based on names. For file names the use of underscores and hyphens to
  indicate two levels of chunking can be very useful:
  \texttt{my-file\_2023-12-25.txt},
  \texttt{some-other-file\_2023-01-02.txt} makes it easy to separate the
  dates from the slugs (the unique identifier for the file), even when
  the slugs have different numbers of components. Requiring students to
  follow file-naming conventions (likely which include some identifier
  for the student) for files that they submit can be a useful way to
  encourage their use of a file-naming system.
\item
  Choose file-naming conventions that sort naturally.

  For dates in file names, we recommend adopting the ISO 8601 standard
  (ISO 2019) (YYYY-MM-DD, for example). This format sorts naturally
  without any special treatment of dates. Padding numbers with 0's so
  that all numbers use the same number of digits also serves this
  purpose, and numbers can be prepended to file names to force a
  particular sorting. (Leaving gaps for future insertions can be helpful
  as well.)
\end{enumerate}

\hypertarget{sec-style}{%
\subsubsection{\texorpdfstring{\textbf{Follow a style guide
consistently.}}{Follow a style guide consistently.}}\label{sec-style}}

Good coding style includes choosing good names for files and variables,
but also includes things like consistent use of whitespace and
indentation; effective use of comments (Spertus 2021); the choice of
data types used for various purposes; and the particular ``coding
dialect'' used (Hadley Wickham 2022; Abouzekry 2012). Consistent use of
whitespace and indentation can be automated in many modern editors and
IDEs (integrated development environments), and we recommend teaching
students to use such tools early in their statistics or data science
programs. Alas, most of the remaining elements of style require
continual human attention.

We recommend adopting a commonly used style guide (like the tidyverse
style guide (Hadley Wickham 2022) or Google's slightly modified version
({``{Google Style Guide}''} 2019) for R, or the PEP 8 style guide for
Python (Rossum, Warsaw, and Coghlan 2023), perhaps with some local
amendments. If a program can adopt a consistent style guide across its
courses, that provides additional advantages and makes things simpler
for students. In any case, adopting and following a style guide is good
both for the improved readability of the resulting code and for practice
in following a style guide.

In R, packages like styler (Müller and Walthert 2022) and formatR (Xie
2022) can assist with style consistency. More generally, modern editors
and IDEs include support for various linters (see, for example, VanTol
(2023), for a discussion of linters for Python) that can detect
violations of code style, inconsistencies, and other coding issues,
sometimes suggesting improvements \emph{as code is being written}.

\hypertarget{create-documents-using-tools-that-support-reproducible-workflows.}{%
\subsubsection{\texorpdfstring{\textbf{Create documents using tools that
support reproducible
workflows.}}{Create documents using tools that support reproducible workflows.}}\label{create-documents-using-tools-that-support-reproducible-workflows.}}

Students familiar with copying and pasting output from one software
application (e.g., Excel) into another (e.g., Word or PowerPoint) should
be encouraged early in their careers to take advantage of other
workflows like R Markdown (Baumer et al. 2014), Quarto (Allaire et al.
2022), and Jupyter notebooks (Granger and Pérez 2021). These tools allow
students to generate multiple document formats (including PDF, Word, and
HTML) from a single source document that contains code in multiple
languages (R, Python, Julia, Observable JavaScript, etc.), the results
of executing that code, and formatted text discussing the process and
results.

These tools provide a convenient way for students to prepare assignments
(and for instructors to prepare learning materials), but more
importantly, they train students to adopt reproducible workflows. Sandve
et al. (2013) offers a set of rules for reproducible computations
research. The first of these is that for each result, it's important to
keep track of where it is produced. A workflow where output or graphics
are copied from one place to another or where manual data wrangling
steps are undertaken outside the documented workflow (Sandve et al.'s
rule 5) obfuscates this important provenance.

Fostering reproducible analyses and workflow early on is valuable for
students, and many instructors who use R emphasize the use of R Markdown
(or more recently, Quarto) starting very early in their introductory
courses (Baumer et al. 2014; Horton et al. 2022). But some practitioners
unfortunately still rely on R scripts that produce auxiliary files that
are then included elsewhere for reporting, perhaps via unautomated
copy-and-paste steps. The use of modern document formats reinforces
proper encapsulation and reproducibility since the documents must be
self-contained and include all information needed to perform the
analysis being presented. Assignments that ask students to repeat an
analysis with an augmented data set or slightly modified task (both
common occurrences in practice), or to undertake workflows which require
separate steps (complicated or time-consuming wrangling followed by
later analysis of time-stamped datasets) can reinforce the power of
these tools.

More advanced students can be taught about parameter-driven documents
(Mahoney 2022) and automated report generation (Beck 2020) and can learn
to create a wider variety of document types, including webpages,
presentations, and dashboards.

\hypertarget{sec-toolkit}{%
\subsubsection{\texorpdfstring{\textbf{Select a coherent, minimal, yet
powerful
toolkit.}}{Select a coherent, minimal, yet powerful toolkit.}}\label{sec-toolkit}}

In most languages, there are many ways to perform some common tasks.
Even when several are equally good on their own merits, selecting a
toolkit consisting of functions that work well together improves
readability and reduces errors that arise from failing to switch from
one standard to another (Çetinkaya-Rundel et al. 2022; Pruim and Horton
2020). The tidyverse (Hadley Wickham et al. 2019) provides one example
of an ``opinionated collection of R packages designed for data science
{[}which{]} share an underlying design philosophy, grammar, and data
structures.'' As such, it provides a good model for the kind of
attention that is needed to produce such a toolkit.

It is especially important that instructors make wise choices about the
toolkit that they present to students. An ideal toolkit should be

\begin{enumerate}
\def\labelenumi{\alph{enumi}.}
\item
  \textbf{Coherent}

  Elements of the toolkit that perform similar tasks should have similar
  structure. This makes the resulting code easier to read and makes it
  easier for students to recall (or even anticipate) code structures.
  Generally speaking, new code written by users should aim to mimic the
  style of and interoperate well with the main elements of the toolkit.
\item
  \textbf{Minimal}

  In most cases, less is more. Learning a few things well, and learning
  how to combine them creatively to perform a wide range of tasks is
  easier and more useful than having a large toolkit of speciality
  functions that each do a specific narrowly-defined task. ``Perfection
  is achieved, not when there is nothing more to add, but when there is
  nothing left to take away.'' (Saint-Exupéry 1984)
\item
  \textbf{Powerful}

  Fighting against the desire to have the toolkit be small and coherent
  is the need to accomplish the tasks at hand (and future tasks as they
  arise). We want our toolkit to be as simple as possible, but no
  simpler.
\end{enumerate}

Each team (or instructor) will need to balance these competing
interests. We advocate against two extremes that are implicit (or
sometimes even explicit) in many coding examples we see: avoiding the
use of standard, well-supported packages (in favor of ``base'' language
constructs), and using code from myriad packages, often with competing
styles and ``mental models''. A well chosen, consistent toolkit both
demonstrates good coding practice and makes it easier for students to
learn the material in a given course. Building a toolkit around a focal
package or collection of packages (like \texttt{tidyverse} or
\texttt{tidymodels} in R or \texttt{numpy} (Harris et al. 2020),
\texttt{pandas} (McKinney et al. 2010) and \texttt{scikit-learn}
(Pedregosa et al. 2011) in Python) can provide a useful filter for
selecting components of a toolkit.

A fortunate recent development is the converging of ideas in R and
Python. This makes it possible to choose a bilingual toolkit with some
level of coherence. Consider the following examples of data wrangling
and plotting in R and in Python, motivated by Hilary Parker's blog post
(Parker 2013) investigating trends in her name's popularity over time.

\begin{Shaded}
\begin{Highlighting}[]
\CommentTok{\# R}
\FunctionTok{library}\NormalTok{(dplyr)}
\FunctionTok{library}\NormalTok{(ggplot2)}
\FunctionTok{library}\NormalTok{(babynames)   }\CommentTok{\# loads babynames data}
\NormalTok{babynames }\SpecialCharTok{|\textgreater{}}
  \FunctionTok{filter}\NormalTok{(name }\SpecialCharTok{\%in\%} \FunctionTok{c}\NormalTok{(}\StringTok{"Hilary"}\NormalTok{, }\StringTok{"Hillary"}\NormalTok{, }\StringTok{"Hilarie"}\NormalTok{, }\StringTok{"Hillarie"}\NormalTok{)) }\SpecialCharTok{|\textgreater{}}
  \FunctionTok{group\_by}\NormalTok{(year, sex) }\SpecialCharTok{|\textgreater{}}
  \FunctionTok{summarise}\NormalTok{(}\AttributeTok{prop =} \FunctionTok{sum}\NormalTok{(prop)) }\SpecialCharTok{|\textgreater{}}
  \FunctionTok{ggplot}\NormalTok{() }\SpecialCharTok{+}
  \FunctionTok{geom\_line}\NormalTok{(}\FunctionTok{aes}\NormalTok{(}\AttributeTok{x =}\NormalTok{ year, }\AttributeTok{y =}\NormalTok{ prop, }\AttributeTok{color =}\NormalTok{ sex)) }\SpecialCharTok{+}
  \FunctionTok{labs}\NormalTok{(}\AttributeTok{title =} \StringTok{"Prevalence of \textquotesingle{}Hilary\textquotesingle{} et al."}\NormalTok{)}
\end{Highlighting}
\end{Shaded}

\begin{figure}[H]

{\centering \includegraphics{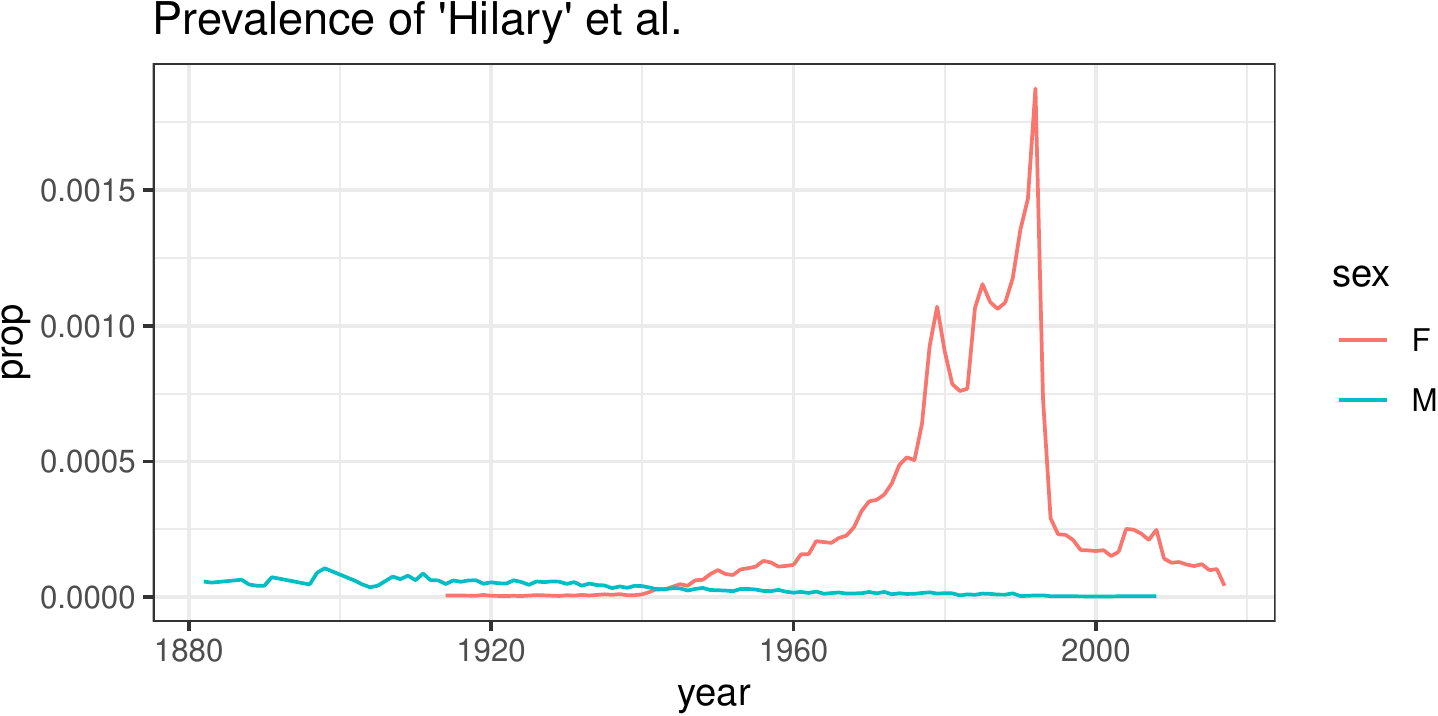}

}

\end{figure}

\hypertarget{annotated-cell-4}{%
\label{annotated-cell-4}}%
\begin{Shaded}
\begin{Highlighting}[]
\CommentTok{\# Python}
\ImportTok{import}\NormalTok{ pandas }\ImportTok{as}\NormalTok{ pd}
\ImportTok{import}\NormalTok{ altair }\ImportTok{as}\NormalTok{ alt}
\ImportTok{from}\NormalTok{ pyreadr }\ImportTok{import}\NormalTok{ read\_r, download\_file}
\NormalTok{url }\OperatorTok{=} \StringTok{"https://github.com/hadley/babynames/raw/master/data/babynames.rda"}
\NormalTok{babynames }\OperatorTok{=}\NormalTok{ read\_r(download\_file(url, }\StringTok{"./babynames.rda"}\NormalTok{))[}\StringTok{"babynames"}\NormalTok{]  }\hspace*{\fill}\NormalTok{\circled{1}}

\NormalTok{(}
\NormalTok{  babynames}
\NormalTok{  .query(}\StringTok{"name in [\textquotesingle{}Hilary\textquotesingle{}, \textquotesingle{}Hillary\textquotesingle{}, \textquotesingle{}Hilarie\textquotesingle{}, \textquotesingle{}Hillarie\textquotesingle{}]"}\NormalTok{)}
\NormalTok{  .groupby(by }\OperatorTok{=}\NormalTok{ [}\StringTok{"year"}\NormalTok{, }\StringTok{"sex"}\NormalTok{], as\_index }\OperatorTok{=} \VariableTok{False}\NormalTok{)  }\hspace*{\fill}\NormalTok{\circled{2}}
\NormalTok{  .aggregate(\{}\StringTok{"prop"}\NormalTok{: }\StringTok{"sum"}\NormalTok{\})}
\NormalTok{  .pipe(alt.Chart, title }\OperatorTok{=} \StringTok{"Prevalence of \textquotesingle{}Hilary\textquotesingle{} et al."}\NormalTok{)}
\NormalTok{  .mark\_line()}
\NormalTok{  .encode(}
\NormalTok{    x }\OperatorTok{=}\NormalTok{ alt.X(}\StringTok{"year"}\NormalTok{).axis(}\BuiltInTok{format} \OperatorTok{=} \StringTok{"4d"}\NormalTok{), }
\NormalTok{    y }\OperatorTok{=} \StringTok{"prop:Q"}\NormalTok{, }
\NormalTok{    color }\OperatorTok{=} \StringTok{"name:N"}\NormalTok{)}
\NormalTok{)}
\end{Highlighting}
\end{Shaded}

\begin{description}
\tightlist
\item[\circled{1}]
Alternatively, the \texttt{reticulate} package provides a way to
communicate data back and forth between R and Python. Using
\texttt{reticulate}, we could access the data (converted from an R data
frame to a pandas data frame) with \texttt{babynames\ =\ r.babynames}.
\item[\circled{2}]
Rossum, Warsaw, and Coghlan (2023) suggests (without much explanation)
\emph{not} putting a space around the assignment \texttt{=} in argument
lists for functions. We choose to include the spaces for better
legibility, especially for projected demonstration code, but we make our
students aware of the more common styling in Python.
\end{description}

\includegraphics[width=0.7\textwidth,height=\textheight]{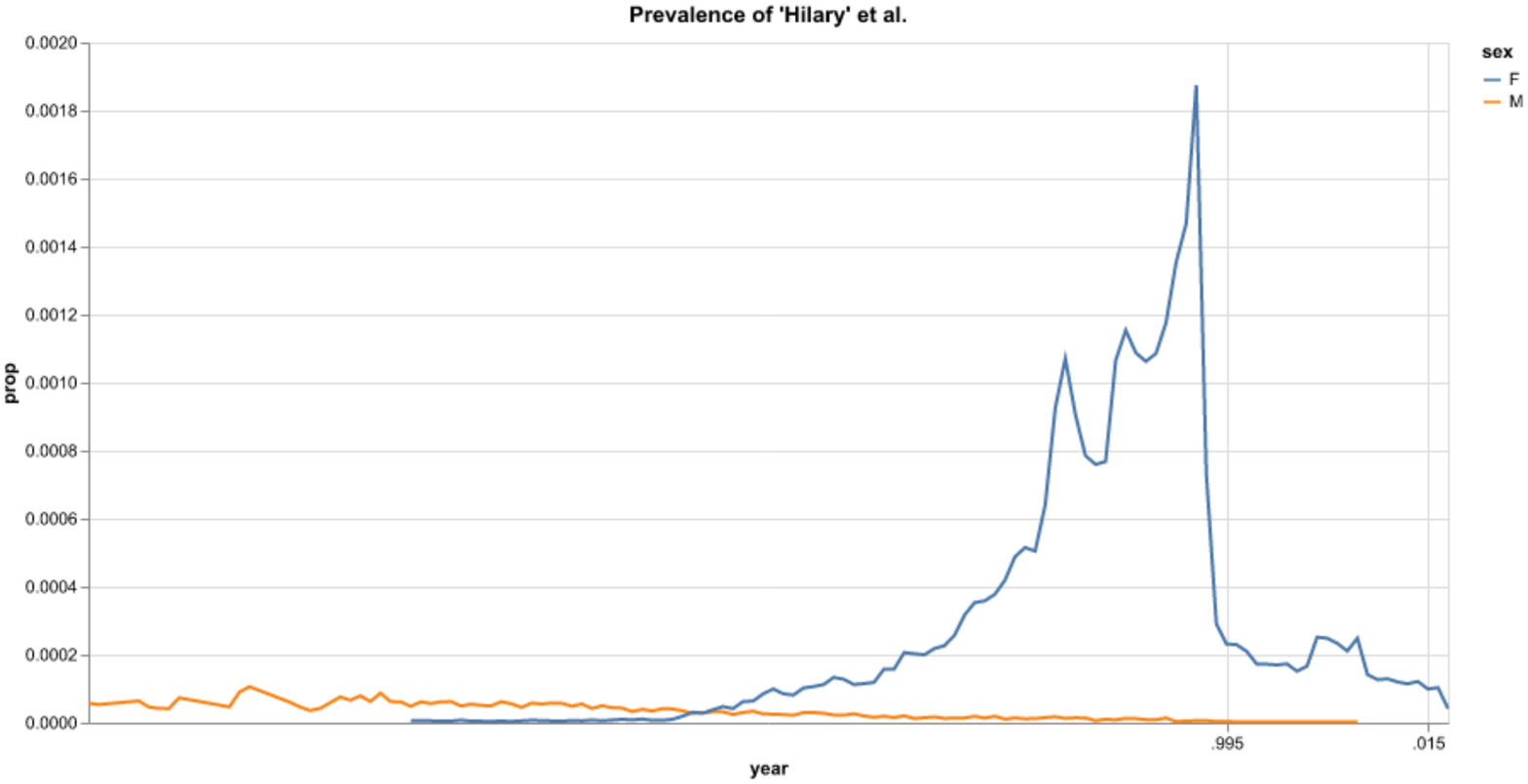}

While some language differences are unavoidable (e.g., method chaining
in Python in place of the pipe (\texttt{\textbar{}\textgreater{}}) in R,
strings in Python where non-standard evaluation can be used to avoid
them in R), and the differences in naming (\texttt{query()}
vs.~\texttt{filter()}, \texttt{aggregate()} vs.~\texttt{summarise()},
etc.) are unfortunate, the basic approach supported by \texttt{dplyr}
and \texttt{pandas} remains quite similar (and based on ideas from SQL).
Similarly, both plotting systems are based on a grammar of graphics
approach.\footnote{Here we use \texttt{altair} (VanderPlas et al. 2018)
  for plotting in Python. \texttt{seaborn.objects}, \texttt{plotly}
  (Inc. 2015) or \texttt{plotnine} (Kibirige et al. 2023) would be other
  Python alternatives based on a grammar of graphics approach (Wilkinson
  et al. 2006). Other options for R include \texttt{ggformula} (Kaplan
  and Pruim, n.d.), \texttt{altair} (Lyttle, Jeppson, and Altair
  Developers 2023), or \texttt{plotly} (Sievert 2020).}

\hypertarget{sec-dry}{%
\subsubsection{\texorpdfstring{\textbf{Don't Repeat Yourself
(DRY).}}{Don't Repeat Yourself (DRY).}}\label{sec-dry}}

Overuse of copy-paste-modify can affect code writing as well as document
creation. This is usually an indication of a bad workflow or poor
encapsulation. Frequent copying and pasting of code may indicate the
need for a function that encapsulates the repeated code into one
location, identifies (via its arguments) what changes, and simplifies
code maintenance, since changes only need to be made in one location
(McConnell 2004). The principle of ``Don't Repeat Yourself'' has been a
mainstay in computer programming at least since Hunt and Thomas (1999)
and serves as a good guideline for statisticians and data scientists as
well.

This is not meant to suggest that scaffolded assignments, where students
are presented with incomplete code to complete, are not useful. We
recognize that many students (and practitioners) often seek to solve
problems by finding working code that does a similar task and modifying
it to obtain the desired result. But students who rely on this as their
sole mechanism for producing code are likely failing to learn important
concepts and structures at a deep level and developing coding habits
that will not serve them (or their colleagues) well. For beginners, this
may require the instructor to modify the toolkit being used or to
provide students with some (but not too many) additional functions
(Pruim and Horton 2020). Before long, however, students should be taught
to write their own simple functions. The authors often do this beginning
from existing code that does one task and asking students to generalize
the task and to create a reusable function to execute it. We illustrate
this using the \texttt{babynames} example from
Section~\ref{sec-toolkit}.

This code can be generalized in several ways. Here we create a function
that allows us to select several names and a range of years.\footnote{We'll
  return to this example in Section~\ref{sec-consistency} to discuss
  ways to make it less brittle.}

\begin{Shaded}
\begin{Highlighting}[]
\CommentTok{\# R}
\NormalTok{babynames\_plot }\OtherTok{\textless{}{-}} 
  \ControlFlowTok{function}\NormalTok{(names, }\AttributeTok{years =} \FunctionTok{c}\NormalTok{(}\DecValTok{1880}\NormalTok{, }\DecValTok{2017}\NormalTok{), }\AttributeTok{data =}\NormalTok{ babynames) \{}
\NormalTok{    data }\SpecialCharTok{|\textgreater{}}
      \FunctionTok{filter}\NormalTok{(year }\SpecialCharTok{\textgreater{}=}\NormalTok{ years[}\DecValTok{1}\NormalTok{], year }\SpecialCharTok{\textless{}=}\NormalTok{ years[}\DecValTok{2}\NormalTok{], name }\SpecialCharTok{\%in\%}\NormalTok{ names) }\SpecialCharTok{|\textgreater{}}
      \FunctionTok{group\_by}\NormalTok{(year, sex, name) }\SpecialCharTok{|\textgreater{}} 
      \FunctionTok{summarise}\NormalTok{(}\AttributeTok{prop =} \FunctionTok{sum}\NormalTok{(prop)) }\SpecialCharTok{|\textgreater{}}
      \FunctionTok{ggplot}\NormalTok{() }\SpecialCharTok{+}
      \FunctionTok{geom\_line}\NormalTok{(}\FunctionTok{aes}\NormalTok{(}\AttributeTok{x =}\NormalTok{ year, }\AttributeTok{y =}\NormalTok{ prop, }\AttributeTok{color =}\NormalTok{ name)) }\SpecialCharTok{+}
      \FunctionTok{facet\_grid}\NormalTok{(sex }\SpecialCharTok{\textasciitilde{}}\NormalTok{ ., }\AttributeTok{scales =} \StringTok{"free"}\NormalTok{) }\SpecialCharTok{+} 
      \FunctionTok{labs}\NormalTok{(}
        \AttributeTok{title =} 
          \FunctionTok{paste0}\NormalTok{(}\StringTok{"Prevalence of "}\NormalTok{, }\FunctionTok{paste}\NormalTok{(names, }\AttributeTok{collapse =} \StringTok{", "}\NormalTok{))) }
\NormalTok{\}}

\FunctionTok{babynames\_plot}\NormalTok{(}\FunctionTok{c}\NormalTok{(}\StringTok{"John"}\NormalTok{, }\StringTok{"Jon"}\NormalTok{, }\StringTok{"Jonathan"}\NormalTok{), }\AttributeTok{years =} \FunctionTok{c}\NormalTok{(}\DecValTok{1950}\NormalTok{, }\DecValTok{2000}\NormalTok{))}
\FunctionTok{babynames\_plot}\NormalTok{(}\FunctionTok{c}\NormalTok{(}\StringTok{"John"}\NormalTok{, }\StringTok{"James"}\NormalTok{, }\StringTok{"Mary"}\NormalTok{), }\AttributeTok{years =} \FunctionTok{c}\NormalTok{(}\DecValTok{1920}\NormalTok{, }\DecValTok{2020}\NormalTok{))}
\end{Highlighting}
\end{Shaded}

\begin{figure}

\begin{minipage}[t]{0.50\linewidth}

{\centering 

\raisebox{-\height}{

\includegraphics{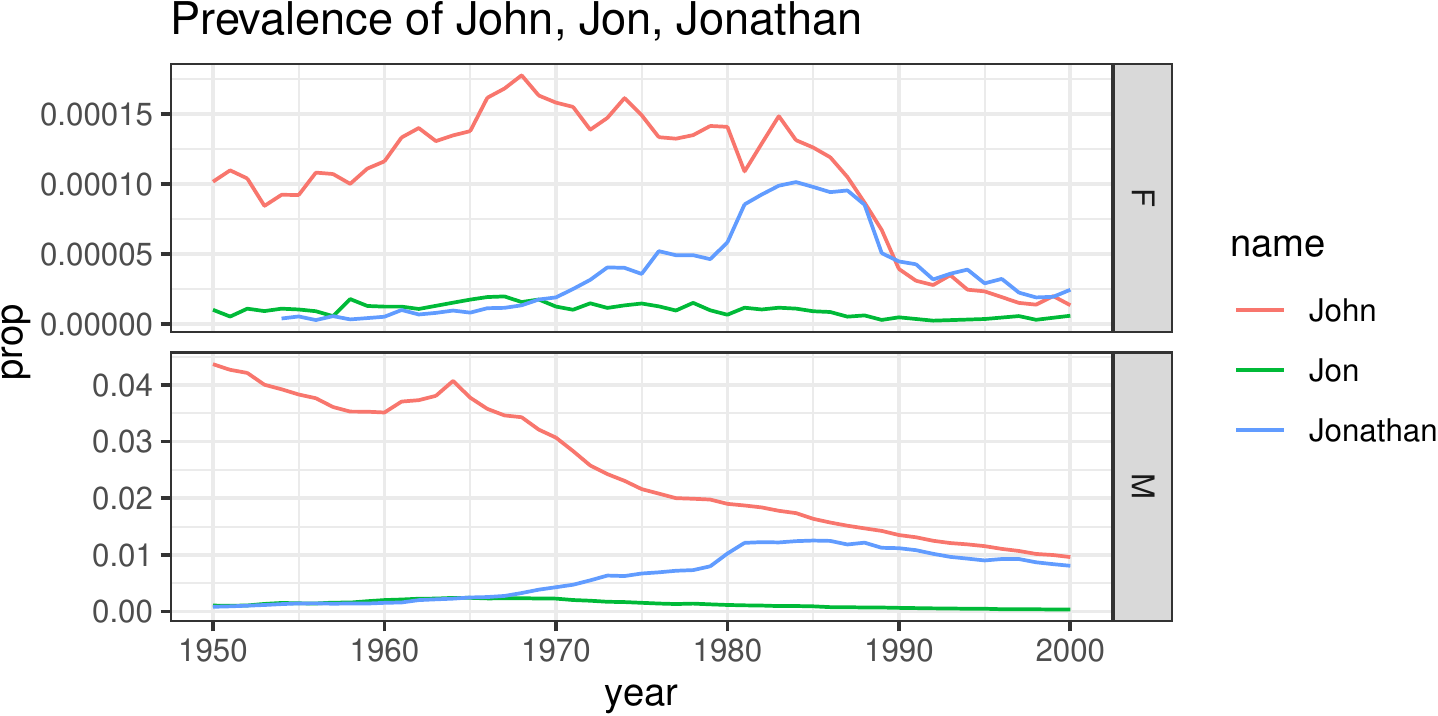}

}

}

\end{minipage}%
\begin{minipage}[t]{0.50\linewidth}

{\centering 

\raisebox{-\height}{

\includegraphics{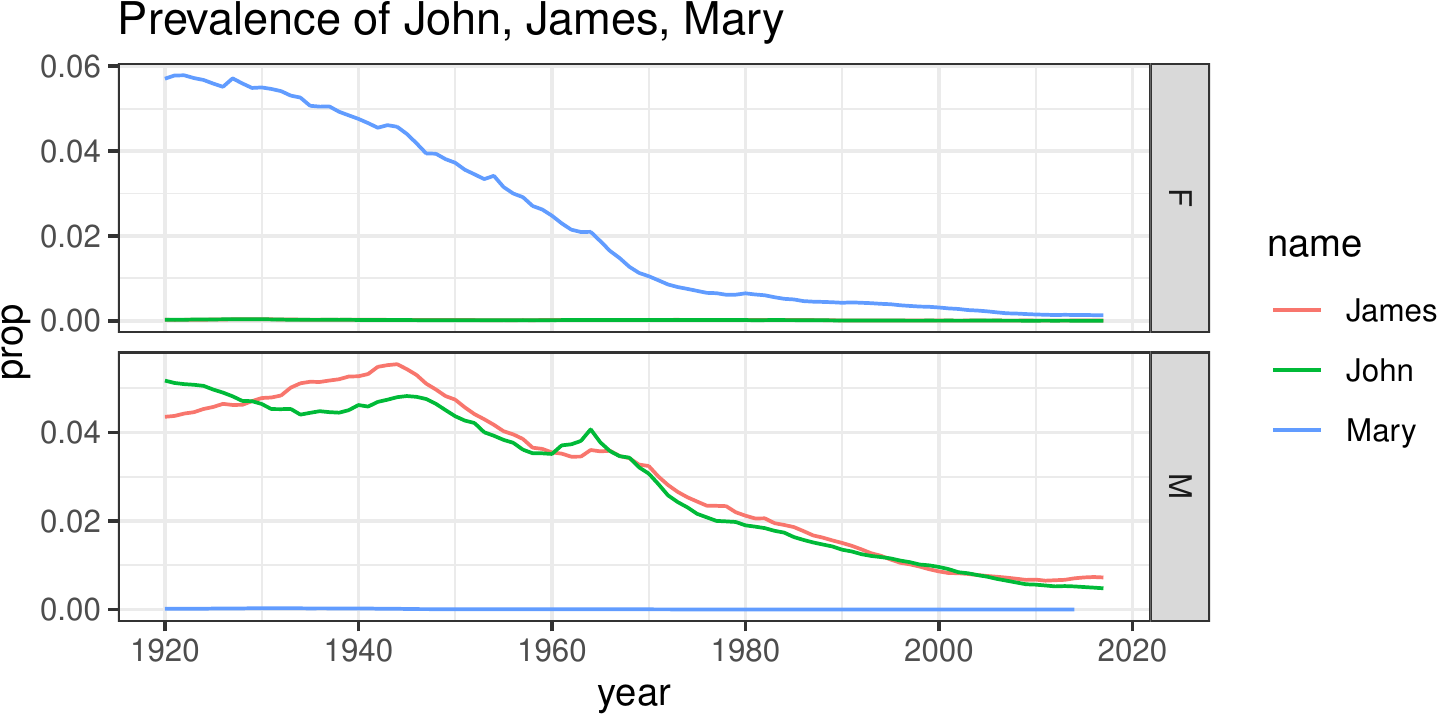}

}

}

\end{minipage}%

\end{figure}

A similar approach can be taken in Python:

\begin{Shaded}
\begin{Highlighting}[]
\CommentTok{\# Python}
\ImportTok{import}\NormalTok{ pandas }\ImportTok{as}\NormalTok{ pd}
\ImportTok{import}\NormalTok{ altair }\ImportTok{as}\NormalTok{ alt}

\KeywordTok{def}\NormalTok{ babynames\_plot(names, years }\OperatorTok{=}\NormalTok{ [}\DecValTok{1880}\NormalTok{, }\DecValTok{2017}\NormalTok{], data }\OperatorTok{=}\NormalTok{ babynames):}
    \ControlFlowTok{return}\NormalTok{ (}
\NormalTok{      data}
\NormalTok{      .query(}\StringTok{"name in @names"}\NormalTok{)}
\NormalTok{      .query(}\StringTok{"year \textgreater{}= @years[0] and year \textless{}= @years[1]"}\NormalTok{)}
\NormalTok{      .pipe(alt.Chart, title }\OperatorTok{=} \StringTok{"Prevalence of "} \OperatorTok{+} \StringTok{", "}\NormalTok{.join(names))}
\NormalTok{      .mark\_line()}
\NormalTok{      .encode(}
\NormalTok{        x }\OperatorTok{=}\NormalTok{ alt.X(}\StringTok{"year"}\NormalTok{).axis(}\BuiltInTok{format}\OperatorTok{=}\StringTok{"4d"}\NormalTok{), }
\NormalTok{        y }\OperatorTok{=} \StringTok{"prop:Q"}\NormalTok{, }
\NormalTok{        color }\OperatorTok{=} \StringTok{"name:N"}\NormalTok{)}
\NormalTok{      .properties(width }\OperatorTok{=} \DecValTok{800}\NormalTok{, height }\OperatorTok{=} \DecValTok{250}\NormalTok{)}
\NormalTok{      .facet(}\StringTok{\textquotesingle{}sex\textquotesingle{}}\NormalTok{, columns }\OperatorTok{=} \DecValTok{1}\NormalTok{)}
\NormalTok{      .resolve\_scale(y }\OperatorTok{=} \StringTok{\textquotesingle{}independent\textquotesingle{}}\NormalTok{)}
\NormalTok{    )}

\NormalTok{babynames\_plot([}\StringTok{"John"}\NormalTok{, }\StringTok{"Jon"}\NormalTok{, }\StringTok{"Jonathan"}\NormalTok{], years }\OperatorTok{=}\NormalTok{ [}\DecValTok{1920}\NormalTok{, }\DecValTok{2020}\NormalTok{])}
\end{Highlighting}
\end{Shaded}

\includegraphics[width=0.7\textwidth,height=\textheight]{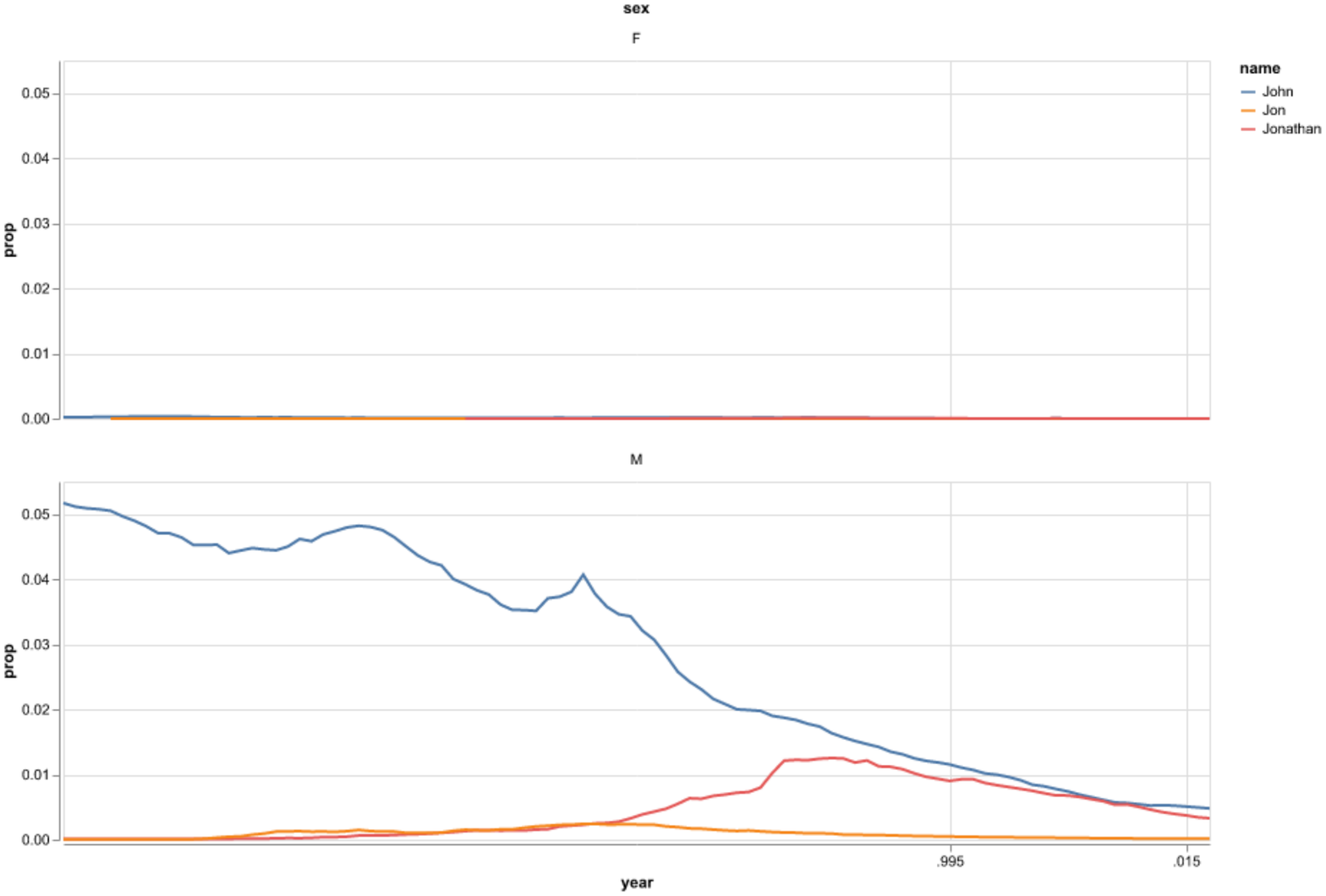}

The ability to write functions also opens up the possibility to use many
general-purpose functions that take functions \emph{as arguments} (see
H. Wickham 2019, chap. 9, and Section~\ref{sec-functional}).

Functions (and data sets) that are useful in multiple projects can be
saved in an R or Python package. Users of R and Python will already be
familiar with packages, since most of the R code that they use comes
from a package. But it is also important to learn how to create
packages, not only to share widely via PyPI, CRAN, bioconductor, or
GitHub, but also for use within ``local'' projects.

H. Wickham (2015) noted that ``R packages are the fundamental units of
reproducible R code.'' Packages provide standard mechanisms for
documentation and testing. We have found that code encapsulated and made
accessible in a package can easily and robustly be included in files
within and across projects. While there are some details to be learned,
the RStudio IDE and the \texttt{devtools} package (Hadley Wickham et al.
2022) take care of many of these details for R users, greatly
simplifying the process of creating a package. As an example, these
tools can be automated to provide feedback methods, e.g.,
\texttt{usethis::use\_github\_action()} to automatically run
\texttt{devtools::check()}.

Students who learn to create packages also gain a firmer understanding
of how packages work, which makes them better users of packages created
by others.

By the end of their programs, data science majors should be facile and
comfortable writing functions and ideally have had experience creating
and maintaining one or more packages.

\hypertarget{sec-functional}{%
\subsubsection{\texorpdfstring{\textbf{Take advantage of a functional
programming
style.}}{Take advantage of a functional programming style.}}\label{sec-functional}}

Even when they are not technically functional programming languages,
many languages -- including R and Python (David-Williams 2023) -- allow
a functional programming style, and this programming style has become
central to many data science workflows.

What is a functional style and why is it important for data science?
Hadley Wickham (H. Wickham 2019) offers this description:

\begin{quote}
It's hard to describe exactly what a functional style is, but generally
I think it means decomposing a big problem into smaller pieces, then
solving each piece with a function or combination of functions. When
using a functional style, you strive to decompose components of the
problem into isolated functions that operate independently. Each
function taken by itself is simple and straightforward to understand;
complexity is handled by composing functions in various ways.
\end{quote}

and this enumeration of benefits:

\begin{quote}
Recently, functional techniques have experienced a surge in interest
because they can produce efficient and elegant solutions to many modern
problems. A functional style tends to create functions that can easily
be analysed in isolation (i.e., using only local information), and hence
is often much easier to automatically optimise or parallelise.
\end{quote}

In R, adopting a functional style also makes it easier to write more
efficient code by taking advantage of many functions written in this
style that bring about a significant boost in performance for many
common tasks. For loops, for example, should mostly be avoided in R and
replaced with equivalent (but more efficient) functional programming
structures. For this reason Burns (2011) describes the use of for loops
as ``speaking R with a C accent -- a strong accent'' (page 17).

The key concept of iterating over an object (a list, a vector, etc.) is
very important, but it is not synonymous with writing for loops. Both R
and Python provide other ways to do this that are more efficient and
more elegant than using for loops. Consider the simple task of summing
the first 100 integers. A student with previous programming experience
was likely taught to approach this task like this:

\begin{Shaded}
\begin{Highlighting}[]
\CommentTok{\# R}
\NormalTok{s }\OtherTok{\textless{}{-}} \DecValTok{0}
\ControlFlowTok{for}\NormalTok{ (i }\ControlFlowTok{in} \DecValTok{1}\SpecialCharTok{:}\DecValTok{100}\NormalTok{) \{}
\NormalTok{  s }\OtherTok{\textless{}{-}}\NormalTok{ s }\SpecialCharTok{+}\NormalTok{ i}
\NormalTok{\}}
\NormalTok{s}
\end{Highlighting}
\end{Shaded}

\begin{verbatim}
[1] 5050
\end{verbatim}

A much more efficient and R-like way to do this is simply:

\begin{Shaded}
\begin{Highlighting}[]
\CommentTok{\# R}
\FunctionTok{sum}\NormalTok{(}\DecValTok{1}\SpecialCharTok{:}\DecValTok{100}\NormalTok{)}
\end{Highlighting}
\end{Shaded}

\begin{verbatim}
[1] 5050
\end{verbatim}

R also includes many functions that are ``vectorized'' so that the
following two lines produce equivalent vectors \texttt{y} containing the
logarithms of each value contained in \texttt{x} (but the second is more
efficient, clearer, and easier to embed in data transformation
workflows).

\begin{Shaded}
\begin{Highlighting}[]
\CommentTok{\# R}
\CommentTok{\# poor use of for loop}
\NormalTok{y }\OtherTok{\textless{}{-}} \FunctionTok{numeric}\NormalTok{(}\FunctionTok{length}\NormalTok{(x)); }\ControlFlowTok{for}\NormalTok{ (i }\ControlFlowTok{in} \DecValTok{1}\SpecialCharTok{:}\FunctionTok{length}\NormalTok{(x)) \{ y[i] }\OtherTok{\textless{}{-}} \FunctionTok{log}\NormalTok{(x[i]) \}}
\CommentTok{\# better}
\NormalTok{y }\OtherTok{\textless{}{-}} \FunctionTok{log}\NormalTok{(x)}
\end{Highlighting}
\end{Shaded}

The \texttt{Vectorize()} function makes it easy for users to create
their own vectorized functions that work in the same way as
\texttt{log()} and many other functions. Additionally, R includes many
functionals (see, for example, H. Wickham 2019, chap. 9) in the
``apply'' family (both in base R and in packages like \texttt{purrr}
(Henry and Wickham 2020)). These functionals provide fine control over
how a function is applied to each item in a list-like structure (or
parallel lists) and how the results are returned. Learning to use these
functions makes code both more efficient and more readable.

Functional programming style, although often not a major component of an
introductory programming course, is useful in many other languages as
well, including Python. (See Kuchling (2023) for an introduction to
functional programming in Python.) The functional programming toolkit
contains a similar collection of tools and concepts, regardless of the
language, so functional programming skills transfer between languages.

The compositional aspect of functional programming combines well with
the use of the pipe operator (\texttt{\textbar{}\textgreater{}}) in R,
which can make sequences of operations easier to read and write. A
similar approach that leans more on an object-oriented implementation in
Python is \textbf{method chaining}. Each method performs a simple task
and returns an object (often of the same type as the original) for which
an additional method can be selected and executed, as was illustrated in
Section~\ref{sec-dry}.

\hypertarget{sec-consistency}{%
\subsubsection{\texorpdfstring{\textbf{Employ consistency
checks.}}{Employ consistency checks.}}\label{sec-consistency}}

Things don't always work. And code written today may be used in the
future in ways that were not anticipated. Avoiding brittle code (like
referring to a column in a data frame by number rather than by name),
building in checks of assumptions, emitting informative messages, and
running automated code tests can reduce the frequency of downstream
errors that may otherwise go undetected. Incorporating such checks can
ensure that code fails safely when there are issues or inconsistencies.

When students are introduced to creating packages, they should learn to
create unit tests. But much earlier, students should be taught to check
their work and their data for obvious problems by answering a few simple
questions like

\begin{itemize}
\tightlist
\item
  Does my data frame have the anticipated shape?
\item
  Do numerical and/or graphical summaries give plausible results?
\item
  Are quantitative features in the expected range (e.g., ages
  non-negative)?
\item
  Have I tried some examples where I know what the correct answer is?
\item
  What happens if I \ldots?
\end{itemize}

Learning to anticipate (and check for) potential problems is a key step
in the progression of a data scientist. Full blown unit testing of the
sort supported by the \texttt{testthat} (Hadley Wickham 2011) or
\texttt{assertr} (Fischetti 2021) packages in R or \texttt{unittest}
({``{unittest --- {Unit} testing framework}''} 2023) or \texttt{pytest}
(Krekel et al. 2004) in Python may not be needed from the start. But it
is good to emphasize the importance of checking for correctness early
on. Data audits are a useful part of any workflow. Visualizations or
tables can help convince us that a data transformation was performed
correctly. Testing code on small examples or examples where the correct
result is known, can help reassure us that the code will work on other
examples. For instructors, the examples that students construct can also
reveal how students are thinking about the task and how their solution
might fail.

From these early efforts at testing, we can build to more robust checks
of correctness. Defensive coding (McNamara and Horton 2018) is an
approach where some investment in runtime checks can help avoid
undetected errors. As a simple example, consider the function we created
in Section~\ref{sec-dry}. Our function is assuming that (at least) two
years are specified and that the first is smaller than the second. It is
safer to test for the expected input type and format and to emit a
helpful error message when something unexpected is provided. Here we
test that one or two numeric values are provided for \texttt{year}.

\begin{Shaded}
\begin{Highlighting}[]
\CommentTok{\# R}
\CommentTok{\# simple defensive coding enforces that years will have two values in}
\CommentTok{\# non{-}decreasing order when we get to rest of function.}
\NormalTok{baby\_plot }\OtherTok{\textless{}{-}} \ControlFlowTok{function}\NormalTok{(names, }\AttributeTok{years =} \FunctionTok{c}\NormalTok{(}\DecValTok{1880}\NormalTok{, }\DecValTok{2017}\NormalTok{)) \{}
  \ControlFlowTok{if}\NormalTok{ (}\SpecialCharTok{!}\FunctionTok{is.numeric}\NormalTok{(years) }\SpecialCharTok{||} \FunctionTok{length}\NormalTok{(years) }\SpecialCharTok{\textless{}} \DecValTok{1} \SpecialCharTok{||} \FunctionTok{length}\NormalTok{(years) }\SpecialCharTok{\textgreater{}} \DecValTok{2}\NormalTok{) \{}
    \FunctionTok{stop}\NormalTok{(}\StringTok{"\textasciigrave{}years\textasciigrave{} should be a numeric vector of length 1 or 2."}\NormalTok{) }
\NormalTok{  \}}
\NormalTok{  years }\OtherTok{\textless{}{-}} \FunctionTok{range}\NormalTok{(years)}
  \CommentTok{\# rest of function}
\NormalTok{\}}
\end{Highlighting}
\end{Shaded}

As another example, if a data analysis is assuming a set of levels for a
categorical variable, these can be confirmed to ensure that data
consistency issues don't arise if a new dataset (perhaps a subset of the
original data set in which some levels do not occur, or an augmented
data set that contains additional levels) is provided. Riederer (2020)
suggests ways to develop robust workflows for data validation.

\hypertarget{sec-debug-help}{%
\subsubsection{\texorpdfstring{\textbf{Learn how to debug and to ask for
help.}}{Learn how to debug and to ask for help.}}\label{sec-debug-help}}

Sometimes we know the code we have written is not working, but what
then? Developing some rudimentary debugging skills, including the use of
a debugger, can make finding and fixing errors much less frustrating and
time consuming. Unfortunately, many instructors do not teach systematic
ways to debug code. The aptly named \emph{What They Forgot to Teach You
About R} (Jennifer Bryan and Hester 2019, chap. 11) includes a chapter
on this important topic specific to R. McConnell (2004) and Thomas and
Hunt (2019) include sections on general principles of debugging (along
with many other useful tips for improving coding practices).

Another vital skill is learning to create a (minimal) reproducible
example (reprex) (Stack Overflow, n.d.). The creation of an example that
clearly and reproducibly demonstrates a problem but doesn't include any
extraneous elements is often the first step toward identifying and
fixing the problem. The \texttt{reprex} package (Jennifer Bryan et al.
2022) in R is useful for making sure that an example is isolated from
things outside the example and makes it easy to share the example with
others.

Reproducible examples can also be useful teaching devices. Consider the
following example which demonstrates some of the differences between
\texttt{as.numeric()} and \texttt{readr::parse\_number()}, each of which
attempts to convert string data to numeric data.

\begin{Shaded}
\begin{Highlighting}[]
\CommentTok{\# R}
\NormalTok{dplyr}\SpecialCharTok{::}\FunctionTok{tibble}\NormalTok{(}
  \AttributeTok{text =} \FunctionTok{c}\NormalTok{(}\StringTok{"5"}\NormalTok{, }\StringTok{"5.3"}\NormalTok{, }\StringTok{"$1.23"}\NormalTok{, }\StringTok{"1,234"}\NormalTok{),}
  \StringTok{\textasciigrave{}}\AttributeTok{as.numeric(text)}\StringTok{\textasciigrave{}} \OtherTok{=} \FunctionTok{as.numeric}\NormalTok{(text),}
  \StringTok{\textasciigrave{}}\AttributeTok{parse\_number(text)}\StringTok{\textasciigrave{}} \OtherTok{=}\NormalTok{ readr}\SpecialCharTok{::}\FunctionTok{parse\_number}\NormalTok{(text)}
\NormalTok{)}
\end{Highlighting}
\end{Shaded}

\begin{verbatim}
# A tibble: 4 x 3
  text  `as.numeric(text)` `parse_number(text)`
  <chr>              <dbl>                <dbl>
1 5                    5                   5   
2 5.3                  5.3                 5.3 
3 $1.23               NA                   1.23
4 1,234               NA                1234   
\end{verbatim}

\hypertarget{sec-github}{%
\subsubsection{\texorpdfstring{\textbf{Get (version) control of the
situation.}}{Get (version) control of the situation.}}\label{sec-github}}

Version control systems (e.g., GitHub) are a key part of a workflow that
fosters many good code practices, including code review (Jenny Bryan and
Hester 2020; Beckman et al. 2021; Fiksel et al. 2019) and distribution
of tasks in a team. Version control tools can also help make software
and analyses more robust (Sandve et al. 2013; Taschuk and Wilson 2017).
Graduates of a data science program should be fluent in the use of at
least one version control system, and programs should give some thought
to where and how this fluency will be developed.

In our experience, students can begin to use basic git commands
(commit/push/pull) fairly early in their programs, perhaps primarily
through an interface included in an IDE like RStudio (RStudio Team 2015)
or Visual Studio Code (Microsoft 2023) or by using GitHub desktop
({``{GitHub Desktop}''} 2023), to avoid the command line interface. But
students need to have explicit instruction later in their programs to
understand what a commit actually is and to learn how to handle a wider
range of situations (effective use of branches, reverting to old
versions, creating and reviewing pull requests, handling merge
conflicts, rebasing, etc.).

The \emph{Learn Git Branching} tutorial (Cottle 2023) provides an
excellent introduction to git with graphical representations of how git
commands affect a repository.

Legacy et al. (2023) describes an approach to teaching undergraduate
students to utilize sophisticated use of version control as part of an
agile data analysis framework.

\hypertarget{sec-language}{%
\subsubsection{\texorpdfstring{\textbf{Be
multilingual.}}{Be multilingual.}}\label{sec-language}}

While proficiency in a language that supports data analysis well is
important for any statistician or data scientist, a willingness to use
other languages and the ability to select an appropriate language for
given task are also important. The use of document formats like R
Markdown, Quarto, or Jupyter notebooks, that support multiple languages
provides an easy way for users more familiar with one language to
incorporate multiple languages in their workflow. The
\texttt{reticulate} (Ushey, Allaire, and Tang 2022) package in R even
provides a simple mechanism for passing data back and forth between R
and Python.

For example, consider the following (somewhat contrived) example, where
data wrangling is done in R, a machine learning model is fit in Python
using Scikit-learn (Pedregosa et al. 2011), and the results are plotted
in R. (The example is contrived because each of these tasks could be
done in either language.) The \texttt{reticulate} package provides an
object \texttt{r} for retrieving data from an R session for use in
Python and an object \texttt{py} for retrieving Python data for use in
R.

\hypertarget{annotated-cell-13}{%
\label{annotated-cell-13}}%
\begin{Shaded}
\begin{Highlighting}[]
\CommentTok{\# R}
\FunctionTok{library}\NormalTok{(reticulate)}
\FunctionTok{library}\NormalTok{(dplyr)}
\FunctionTok{library}\NormalTok{(tidyr)}
\FunctionTok{library}\NormalTok{(palmerpenguins)}
\NormalTok{penguins }\OtherTok{\textless{}{-}} 
\NormalTok{  penguins }\SpecialCharTok{|\textgreater{}}
  \FunctionTok{filter}\NormalTok{(island }\SpecialCharTok{==} \StringTok{"Dream"}\NormalTok{) }\SpecialCharTok{|\textgreater{}}
  \FunctionTok{mutate}\NormalTok{(}
    \AttributeTok{species =} \FunctionTok{case\_when}\NormalTok{(    }\hspace*{\fill}\NormalTok{\circled{1}}
\NormalTok{      species }\SpecialCharTok{==} \StringTok{"Adelie"} \SpecialCharTok{\textasciitilde{}} \DecValTok{0}\NormalTok{,}
\NormalTok{      species }\SpecialCharTok{==} \StringTok{"Chinstrap"} \SpecialCharTok{\textasciitilde{}} \DecValTok{1}
\NormalTok{    )}
\NormalTok{  ) }\SpecialCharTok{|\textgreater{}}
  \FunctionTok{drop\_na}\NormalTok{()}

\NormalTok{X\_penguins }\OtherTok{\textless{}{-}} \FunctionTok{model.matrix}\NormalTok{(species }\SpecialCharTok{\textasciitilde{}} \SpecialCharTok{{-}}\DecValTok{1} \SpecialCharTok{+}\NormalTok{ ., }\AttributeTok{data =}\NormalTok{ penguins)  }\hspace*{\fill}\NormalTok{\circled{2}}
\NormalTok{y\_penguins }\OtherTok{\textless{}{-}}\NormalTok{ penguins[[}\StringTok{"species"}\NormalTok{]]}
\end{Highlighting}
\end{Shaded}

\begin{description}
\tightlist
\item[\circled{1}]
This recodes species as 0/1 for use in Python ML algorithm shortly.
\item[\circled{2}]
\texttt{model.matrix()} recodes categorical data using indicator
variables.
\end{description}

\hypertarget{annotated-cell-14}{%
\label{annotated-cell-14}}%
\begin{Shaded}
\begin{Highlighting}[]
\CommentTok{\# Python}
\ImportTok{import}\NormalTok{ numpy }\ImportTok{as}\NormalTok{ np}
\ImportTok{from}\NormalTok{ sklearn.svm }\ImportTok{import}\NormalTok{ SVC    }\hspace*{\fill}\NormalTok{\circled{1}}
\ImportTok{from}\NormalTok{ sklearn.preprocessing }\ImportTok{import}\NormalTok{ StandardScaler   }\hspace*{\fill}\NormalTok{\circled{2}}
\ImportTok{from}\NormalTok{ sklearn.model\_selection }\ImportTok{import}\NormalTok{ train\_test\_split}
\ImportTok{from}\NormalTok{ sklearn.pipeline }\ImportTok{import}\NormalTok{ Pipeline}

\NormalTok{X\_penguins }\OperatorTok{=}\NormalTok{ r.X\_penguins }
\NormalTok{y\_penguins }\OperatorTok{=}\NormalTok{ np.array(r.y\_penguins)}
\NormalTok{indices }\OperatorTok{=}\NormalTok{ np.arange(X\_penguins.shape[}\DecValTok{0}\NormalTok{])}
\NormalTok{penguins\_split }\OperatorTok{=}\NormalTok{ train\_test\_split(}
\NormalTok{  X\_penguins, y\_penguins, indices, }
\NormalTok{  test\_size }\OperatorTok{=} \FloatTok{0.3}\NormalTok{, random\_state }\OperatorTok{=} \DecValTok{0}
\NormalTok{  )}
\NormalTok{X\_train, X\_test, y\_train, y\_test, indices\_train, indices\_test }\OperatorTok{=}\NormalTok{ penguins\_split}
\NormalTok{pipe }\OperatorTok{=}\NormalTok{ Pipeline([(}\StringTok{"scaler"}\NormalTok{, StandardScaler()), (}\StringTok{"svc"}\NormalTok{, SVC())])}
\NormalTok{pipe.fit(X\_train, y\_train)}
\NormalTok{pipe.score(X\_train, y\_train)}
\NormalTok{pipe.score(X\_test, y\_test)}
\NormalTok{predictions }\OperatorTok{=}\NormalTok{ pipe.predict(X\_penguins)}
\end{Highlighting}
\end{Shaded}

\begin{description}
\tightlist
\item[\circled{1}]
\texttt{SVC} = Support Vector Classifier
\item[\circled{2}]
\texttt{StandardScalar} transforms to have mean = 0; sd = 1
\end{description}

\begin{verbatim}
Pipeline(steps=[('scaler', StandardScaler()), ('svc', SVC())])
1.0
0.9459459459459459
\end{verbatim}

As a final note, we mention the usefulness of learning the basics of
shell languages such as BASH and common command line utilities,
something often entirely unfamiliar to our students. Familiarity with
basic shell commands to create directories and manage files can
eventually be leveraged for scripting and ``batch'' processing, taking
advantage of additional utility functions.

\hypertarget{summary}{%
\subsubsection{Summary}\label{summary}}

We conclude this section with a succinct enumeration of our top ten
list.

\begin{enumerate}
\def\labelenumi{\arabic{enumi}.}
\tightlist
\item
  Choose good names.
\item
  Follow a style guide consistently.
\item
  Create documents using tools that support reproducible workflows.
\item
  Select a coherent, minimal, yet powerful toolkit.
\item
  Don't Repeat Yourself (DRY).
\item
  Take advantage of a functional programming style.
\item
  Employ consistency checks.
\item
  Learn how to debug and to ask for help.
\item
  Get (version) control of the situation.
\item
  Be multilingual.
\end{enumerate}

This list is intended to be illustrative rather than exhaustive. Each is
actionable and teachable, and together they help achieve the four C's
from Section~\ref{sec-fourc} (correctness, clarity, containment, and
consistency) that typify high quality code.

\hypertarget{discussion}{%
\section{Discussion}\label{discussion}}

Improved coding practices are vital to good statistics and data science.
Our past practices may have been ``good enough'' but the increasing
complexity of analyses and the need to address increasingly
sophisticated questions requires us to up our game. A 2019 National
Academies report called for educational institutions, professional
societies, researchers, and funders to work to improve computational
reproducibility (National Academies of Science, Engineering, and
Medicine 2019). Adopting good coding practices is one part of addressing
this need. Defining and explaining the concept of code quality is
therefore a challenge faced by educators (Borstler et al. 2017) and
industry alike. Unfortunately, typical reward structures do not pay
sufficient attention to code quality or other aspects of responsible
analysis.

There are open questions regarding when and how these practices should
be included in the statistics and data science curriculum. The National
Academies of Science, Engineering, and Medicine (2018) report indicated
that data acumen requires multiple opportunities to engage in the entire
data analysis cycle. No matter how much we teach, students must be given
the time to learn and to consolidate their knowledge into their habits
and workflow.

Instructors play an important role (Keuning, Heeren, and Jeuring 2019;
Theobold, Hancock, and Mannheimer 2021). In academic programs it is
important to begin establishing these practices early, to reinforce them
often, and to expect students to adopt more and more habits of good
programmers as they progress through their programs of study.

Based on our experience, we offer the following advice to instructors
seeking to improve the quality of the code their students produce.

\begin{enumerate}
\def\labelenumi{\arabic{enumi}.}
\item
  \textbf{Hold yourself (as instructor) to a higher standard while
  gently guiding students to better coding practices.}

  In reference to internet protocols, Postel (1980) coined the phrase
  ``be conservative in what you send out and liberal in what you
  accept'': this philosophy seems equally appropriate when teaching or
  mentoring coding practices. A good music teacher will always
  demonstrate good technique and musicianship, even if the student is
  not capable of performing at the same level, and perhaps not even able
  to appreciate some aspects of the teacher's playing. The same should
  be true of the code that instructors present to students. Dogucu and
  Çetinkaya-Rundel (2022) discuss the importance of such role modeling
  in the context of teaching reproducibility.

  Instructors should seek opportunities to improve their own coding
  practices as well as the practices of their students. Collaborating
  with colleagues to review code examples and exchange ideas (and
  increase consistency throughout the program) is a good place to start.
  Seeking out authors who exemplify good code, and using or imitating
  their code is also helpful. Making code publicly available and
  participating in community supported open source projects are
  additional ways to improve one's coding practices.
\item
  \textbf{Start small and adopt a progressive approach} that provides
  ample opportunities to practice and gradually becomes more strict.

  In addition to modeling good technique and musicianship, a good music
  teacher also focuses teaching attention on those areas where the
  student can benefit most. This assessment takes into account both the
  immediate reward for the student (through improvement that the student
  can readily appreciate) and long-term goals (e.g., avoiding or
  breaking a bad habit early that will be harder to break later, even if
  the student doesn't yet understand the importance of the particular
  habit). Once this assessment is complete, the music teacher must
  select an appropriate set of exercises, etudes, and pieces that both
  isolate particular skills and help incorporate them into the way they
  routinely practice and perform -- and encourage the student to
  practice, practice, practice.

  The task of a data science instructor is similar. It's important to
  distinguish which details matter a lot and which are more minor. Over
  the course of a multi-year program, a progressive approach that
  expects higher quality code from students in each subsequent course
  should help students to endeavor to write good code as a matter of
  course and appreciate the importance of doing so.
\item
  \textbf{Use live coding demonstrations to model appropriate practice}.

  Talk through your own coding, debugging, and analysis process as you
  code so students can hear your thought process. It's valuable to get
  the students involved, having them make code suggestions, or suggest
  improvements to your code. Hadley Wickham (2018) provides an excellent
  example of an expert analyst demonstrating their workflow and process
  with a growth mindset (which includes making multiple errors and
  corrections throughout the process).
\item
  \textbf{Regularly comment on student code practices} -- good and bad
  -- and include these in assessment rubrics.

  Students will often search the internet for help and start using code
  that is nothing like what has been demonstrated in class. Beginners
  are typically unable to bring the two into alignment, if that is even
  possible. If we value good coding practices, then we must assess them
  and provide formative feedback so that students understand what is
  valued and know when they are making (or in need of making) progress.
  Stegeman, Barendsen, and Smetsers (2016) provide some useful guidance
  about assessment of code quality in the context of introductory
  programming courses, much of which applies to data science courses as
  well.
\item
  \textbf{Ask students to address code issues before providing help on
  debugging}.

  Code issues like styling, formatting, naming can be improved before
  the main debugging issue is addressed. Not only does this make it
  easier for the instructor to read and understand the student's code,
  but the process of improving the code may lead the student to discover
  for themselves what isn't working.
\item
  \textbf{Provide opportunities for students to collaborate on code and
  to refactor their own code.}

  Giving students the opportunity to read other students' code and to
  see how their classmates respond to their own code can reinforce the
  importance of using good coding practices. Code revision (e.g., Jenny
  Bryan 2018) and improvement is just as important for developing good
  coding practices as multiple drafts are in a composition course.
\end{enumerate}

Many practical challenges remain. Consider approaches to naming. The
highly heterogeneous practices used with the R and Python communities
doesn't do us any favors in this realm. Consider the following set of
base R functions as one example: \texttt{row.names()},
\texttt{rownames()} (but \texttt{row.names()} is preferred),
\texttt{colnames()} (but no \texttt{col.names()}), \texttt{colSums()}.
There is no apparent method to the maddening naming choices here.

This diversity of style conventions makes it hard for users and models
poor coding practice. In this context it is even more important for
analysts and instructors to make consistent, clear naming choices and to
establish and follow a style guideline .

To close, we quote Peter Norvig (personal communication), who noted the
critical importance of a ``meticulous attention to detail'' in data
scientists. Many job descriptions also feature this as an attribute of
successful job candidates. We believe that such attention to detail will
be easier to inculcate within a structure that rewards building better
code.

\hypertarget{acknowledgments}{%
\section{Acknowledgments}\label{acknowledgments}}

Thanks to NSF IISE award 1923388 (DSC-WAV) for partial support of this
project.

\hypertarget{references}{%
\section*{References}\label{references}}
\addcontentsline{toc}{section}{References}

\hypertarget{refs}{}
\begin{CSLReferences}{1}{0}
\leavevmode\vadjust pre{\hypertarget{ref-Abouzekry:2012}{}}%
Abouzekry, Abdullah. 2012. {``10 Tips for Better Coding.''}
\url{https://www.sitepoint.com/10-tips-for-better-coding}.

\leavevmode\vadjust pre{\hypertarget{ref-quarto}{}}%
Allaire, J. J., Charles Teague, Carlos Scheidegger, Yihui Xie, and
Christophe Dervieux. 2022. {``{Quarto}.''}
\url{https://doi.org/10.5281/zenodo.5960048}.

\leavevmode\vadjust pre{\hypertarget{ref-Aruliah:2012}{}}%
Aruliah, Dhavide A., C. Titus Brown, Neil P. Chue Hong, Matt Davis,
Richard T. Guy, Steven H. D. Haddock, Katy Huff, et al. 2012. {``Best
Practices for Scientific Computing.''} \emph{PLOS Biology} 12.
\url{https://doi.org/10.1371/journal.pbio.1001745}.

\leavevmode\vadjust pre{\hypertarget{ref-Augspurger:2016}{}}%
Augspurger, Tom. 2016. {``Modern Pandas (Part 2): Method Chaining.''}
\url{https://tomaugspurger.github.io/posts/method-chaining/\#method-chaining}.

\leavevmode\vadjust pre{\hypertarget{ref-Ball:2022}{}}%
Ball, Richard, Norm Medeiros, Nicholas W. Bussberg, and Aneta Piekut.
2022. {``An Invitation to Teaching Reproducible Research: {L}essons from
a Symposium.''} \emph{Journal of Statistics and Data Science Education},
1--10. \url{https://doi.org/10.1080/26939169.2022.2099489}.

\leavevmode\vadjust pre{\hypertarget{ref-Baumer:RMarkdown}{}}%
Baumer, B. S., M. Çetinkaya-Rundel, A. Bray, L. Loi, and N. J. Horton.
2014. {``R {M}arkdown: {I}ntegrating a Reproducible Analysis Tool into
Introductory Statistics.''} \emph{Technology Innovations in Statistics
Education} 8 (1). \url{https://escholarship.org/uc/item/90b2f5xh}.

\leavevmode\vadjust pre{\hypertarget{ref-Beck:2020}{}}%
Beck, Marcus. 2020. {``Automated Reporting in {Tampa Bay} with Open
Science.''}
\url{https://www.openscapes.org/blog/2020/11/16/tampa-bay-reporting/}.

\leavevmode\vadjust pre{\hypertarget{ref-Beckman:2021}{}}%
Beckman, Matthew D., Mine Çetinkaya-Rundel, Nicholas J. Horton, Colin W.
Rundel, Adam J. Sullivan, and Maria Tackett. 2021. {``Implementing
Version Control with {Git} and {GitHub} as a Learning Objective in
Statistics and Data Science Courses.''} \emph{Journal of Statistics and
Data Science Education} 29 (1): 1--35.
\url{https://doi.org/10.1080/10691898.2020.1848485}.

\leavevmode\vadjust pre{\hypertarget{ref-Borstler:2017}{}}%
Borstler, Jürgen, Harald Störrle, Daniel Toll, Jelle van Assema, Rodrigo
Duran, Sara Hooshangi, Johan Jeuring, Hieke Keuning, Carsten Kleiner,
and Bonnie MacKeller. 2017. {``I Know It When {I} See It: {P}erceptions
of Code Quality {ITiCSE} '17 Working Group Report.''}
\url{https://dl.acm.org/doi/abs/10.1145/3174781.3174785}.

\leavevmode\vadjust pre{\hypertarget{ref-Bryan:WTF}{}}%
Bryan, Jennifer, and Jim Hester. 2019. {``What They Forgot to Teach You
about {R}.''} \url{https://rstats.wtf/}.

\leavevmode\vadjust pre{\hypertarget{ref-reprex}{}}%
Bryan, Jennifer, Jim Hester, David Robinson, Hadley Wickham, and
Christophe Dervieux. 2022. \emph{{reprex}: {P}repare Reproducible
Example Code via the Clipboard}.
\url{https://CRAN.R-project.org/package=reprex}.

\leavevmode\vadjust pre{\hypertarget{ref-Bryan:naming}{}}%
Bryan, Jenny. 2015. {``Naming Things.''}
\url{http://www2.stat.duke.edu/~rcs46/lectures_2015/01-markdown-git/slides/naming-slides/naming-slides.pdf}.

\leavevmode\vadjust pre{\hypertarget{ref-Bryan:2018}{}}%
---------. 2018. {``Code Smells and Feels.''}
\url{https://www.youtube.com/watch?v=7oyiPBjLAWY}.

\leavevmode\vadjust pre{\hypertarget{ref-Bryan:HappyGit}{}}%
Bryan, Jenny, and Jim Hester. 2020. \emph{Happy Git and GitHub for the
{useR}}. \url{https://happygitwithr.com}.

\leavevmode\vadjust pre{\hypertarget{ref-Burns:inferno}{}}%
Burns, Patrick. 2011. \emph{The {R} {I}nferno}.
\url{https://www.burns-stat.com/pages/Tutor/R/_inferno.pdf}.

\leavevmode\vadjust pre{\hypertarget{ref-Carey:2018}{}}%
Carey, Maureen A., and Jason A. Papin. 2018. {``Ten Simple Rules for
Biologists Learning to Program.''} \emph{PLOS Computational Biology} 14
(1). \url{https://doi.org/10.1371/journal.pcbi.1005871}.

\leavevmode\vadjust pre{\hypertarget{ref-Cetinkaya-Rundel:2022}{}}%
Çetinkaya-Rundel, Mine, Johanna Hardin, Benjamin Baumer, Amelia
McNamara, Nicholas J. Horton, and Colin Rundel. 2022. {``An Educator's
Perspective of the Tidyverse.''} \emph{Technology Innovations in
Statistics Education} 14 (1). \url{https://doi.org/10.5070/T514154352}.

\leavevmode\vadjust pre{\hypertarget{ref-learn-git-branching}{}}%
Cottle, Peter. 2023. {``{Learn Git Branching}.''} website.
\url{https://learngitbranching.js.org}.

\leavevmode\vadjust pre{\hypertarget{ref-DanaCenter}{}}%
Dana Center. 2021. {``Data Science Course Framework.''} 2021.
\url{https://www.utdanacenter.org/sites/default/files/2021-05/data/_science/_course/_framework/_2021/_final.pdf}.

\leavevmode\vadjust pre{\hypertarget{ref-David-Williams:2023}{}}%
David-Williams, Stephen. 2023. {``{Functional Programming in Data
Engineering with Python {\ifmmode---\else\textemdash\fi} Part 1}.''}
\emph{Medium}, June.
\url{https://medium.com/data-engineer-things/functional-programming-in-data-engineering-with-python-part-1-c2c4f677f749}.

\leavevmode\vadjust pre{\hypertarget{ref-Dogucu:2022}{}}%
Dogucu, Mine, and Mine Çetinkaya-Rundel. 2022. {``Tools and
Recommendations for Reproducible Teaching.''} \emph{In Revision}.
\url{https://arxiv.org/pdf/2103.12793}.

\leavevmode\vadjust pre{\hypertarget{ref-Fiksel:2019}{}}%
Fiksel, Jacob, Leah R. Jager, Johanna S. Hardin, and Margaret A. Taub.
2019. {``Using {GitHub Classroom} to Teach Statistics.''} \emph{Journal
of Statistics Education} 27 (2): 110--19.
\url{https://doi.org/10.1080/10691898.2019.1617089}.

\leavevmode\vadjust pre{\hypertarget{ref-Filazzola:2022}{}}%
Filazzola, Alessandro, and CJ Lortie. 2022. {``A Call for Clean Code to
Effectively Communicate Science.''} \emph{Methods in Ecology and
Evolution}.
\url{https://besjournals.onlinelibrary.wiley.com/doi/pdfdirect/10.1111/2041-210X.13961}.

\leavevmode\vadjust pre{\hypertarget{ref-assertr}{}}%
Fischetti, Tony. 2021. \emph{{assertr}: {A}ssertive Programming for r
Analysis Pipelines}. \url{https://CRAN.R-project.org/package=assertr}.

\leavevmode\vadjust pre{\hypertarget{ref-Fowler:TwoHardThings}{}}%
Fowler, Martin. 2009. {``{TwoHardThings}.''}
\url{https://www.martinfowler.com/bliki/TwoHardThings.html}.

\leavevmode\vadjust pre{\hypertarget{ref-SocialScience:2022}{}}%
Gentzkow, Matthew, and Jesse M. Shapiro. 2022. {``Code and Data for the
Social Sciences: {A} Practitioner's Guide.''}
\url{https://web.stanford.edu/~gentzkow/research/CodeAndData.xhtml}.

\leavevmode\vadjust pre{\hypertarget{ref-atlassian}{}}%
Ghani, Usman. 2022. {``4 Tips to Improve Code Quality.''}
\url{https://www.atlassian.com/blog/add-ons/4-tips-to-improve-code-quality}.

\leavevmode\vadjust pre{\hypertarget{ref-github-desktop}{}}%
{``{GitHub Desktop}.''} 2023. \emph{GitHub Desktop}. website.
\url{https://desktop.github.com}.

\leavevmode\vadjust pre{\hypertarget{ref-google-style}{}}%
{``{Google Style Guide}.''} 2019.
\url{https://google.github.io/styleguide/Rguide.html}.

\leavevmode\vadjust pre{\hypertarget{ref-Granger:2021}{}}%
Granger, Brian E., and Fernando Pérez. 2021. {``Jupyter: Thinking and
Storytelling with Code and Data.''} \emph{Computing in Science \&
Engineering} 23 (2): 7--14.
\url{https://doi.org/10.1109/MCSE.2021.3059263}.

\leavevmode\vadjust pre{\hypertarget{ref-Hardin:2021}{}}%
Hardin, Johanna, Nicholas J. Horton, Deborah Nolan, and Duncan Temple
Lang. 2021. {``Computing in the Statistics Curricula: {A} 10-Year
Retrospective.''} \emph{Journal of Statistics and Data Science
Education} 29 (sup1): S4--6.
\url{https://doi.org/10.1080/10691898.2020.1862609}.

\leavevmode\vadjust pre{\hypertarget{ref-numpy}{}}%
Harris, Charles R., K. Jarrod Millman, Stéfan J van der Walt, Ralf
Gommers, Pauli Virtanen, David Cournapeau, Eric Wieser, et al. 2020.
{``Array Programming with {NumPy}.''} \emph{Nature} 585: 357--62.
\url{https://doi.org/10.1038/s41586-020-2649-2}.

\leavevmode\vadjust pre{\hypertarget{ref-purrr}{}}%
Henry, Lionel, and Hadley Wickham. 2020. \emph{{purrr}: {F}unctional
Programming Tools}. \url{https://CRAN.R-project.org/package=purrr}.

\leavevmode\vadjust pre{\hypertarget{ref-Horton:2022}{}}%
Horton, Nicholas J., Rohan Alexander, Micaela Parker, Aneta Piekut, and
Colin Rundel. 2022. {``The Growing Importance of Reproducibility and
Responsible Workflow in the Data Science and Statistics Curriculum.''}
\emph{Journal of Statistics and Data Science Education} 30 (3): 207--8.
\url{https://doi.org/10.1080/26939169.2022.2141001}.

\leavevmode\vadjust pre{\hypertarget{ref-Hunt:1999}{}}%
Hunt, Andrew, and David Thomas. 1999. \emph{The Pragmatic Programmer}.
Boston, MA: Addison Wesley.

\leavevmode\vadjust pre{\hypertarget{ref-plotly}{}}%
Inc., Plotly Technologies. 2015. {``Collaborative Data Science.''}
Montreal, QC: Plotly Technologies Inc. 2015. \url{https://plot.ly}.

\leavevmode\vadjust pre{\hypertarget{ref-iso-8601}{}}%
ISO. 2019. {``{ISO} 8601 --- Date and Time Format.''}
\url{https://www.iso.org/iso-8601-date-and-time-format.html}.

\leavevmode\vadjust pre{\hypertarget{ref-ggformula}{}}%
Kaplan, Daniel, and Randall Pruim. n.d. \emph{Ggformula: {F}ormula
Interface to the Grammar of Graphics}.
\url{https://github.com/ProjectMOSAIC/ggformula}.

\leavevmode\vadjust pre{\hypertarget{ref-Keuning:2017}{}}%
Keuning, Hieke, Bastiaan Heeren, and Johan Jeuring. 2017. {``Code
Quality Issues in Student Programs.''}
\url{https://dl.acm.org/doi/10.1145/3059009.3059061}.

\leavevmode\vadjust pre{\hypertarget{ref-Keuning:2019}{}}%
---------. 2019. {``How Teachers Would Help Students to Improve Their
Code.''} \url{https://dl.acm.org/doi/10.1145/3304221.3319780}.

\leavevmode\vadjust pre{\hypertarget{ref-plotnine}{}}%
Kibirige, Hassan, Greg Lamp, Jan Katins, gdowding, austin, Florian
Finkernagel, matthias-k, et al. 2023. {``Has2k1/Plotnine: V0.12.1.''}
Zenodo. \url{https://doi.org/10.5281/zenodo.8171350}.

\leavevmode\vadjust pre{\hypertarget{ref-pytest}{}}%
Krekel, Holger, Bruno Oliveira, Ronny Pfannschmidt, Floris Bruynooghe,
Brianna Laugher, and Florian Bruhin. 2004. {``Pytest 7.1.''}
\url{https://github.com/pytest-dev/pytest}.

\leavevmode\vadjust pre{\hypertarget{ref-functional-python}{}}%
Kuchling, A. M. 2023. {``{Functional Programming HOWTO}.''} \emph{Python
Documentation}. \url{https://docs.python.org/3/howto/functional.html}.

\leavevmode\vadjust pre{\hypertarget{ref-Legacy:2023}{}}%
Legacy, Chelsey, Andrew Zieffler, Benjamin S. Baumer, Valerie Barr, and
Nicholas J. Horton. 2023. {``Facilitating Team-Based Data Science:
Lessons Learned from the DSC-WAV Project.''} \emph{Foundations of Data
Science} 5 (2): 244--65. \url{https://doi.org/10.3934/fods.2022003}.

\leavevmode\vadjust pre{\hypertarget{ref-Lyman:2021}{}}%
Lyman, Isaac. 2021. {``Code Quality: {A} Concern for Businesses, Bottom
Lines, and Empathetic Programmers.''}
\url{https://stackoverflow.blog/2021/10/18/code-quality-a-concern-for-businesses-bottom-lines-and-empathetic-programmers}.

\leavevmode\vadjust pre{\hypertarget{ref-altair-r}{}}%
Lyttle, Ian, Haley Jeppson, and Altair Developers. 2023. \emph{{a}ltair:
{I}nterface to {'Altair'}}.
\url{https://CRAN.R-project.org/package=altair}.

\leavevmode\vadjust pre{\hypertarget{ref-Mahoney:online}{}}%
Mahoney, Mike. 2022. {``How to Use Quarto for Parameterized
Reporting.''}
\url{https://www.mm218.dev/posts/2022-08-04-how-to-use-quarto-for-parameterized-reporting}.

\leavevmode\vadjust pre{\hypertarget{ref-McConnel:Code_Complete}{}}%
McConnell, Steve. 2004. \emph{Code Complete}. 2nd ed. Cisco Press.

\leavevmode\vadjust pre{\hypertarget{ref-pandas}{}}%
McKinney, Wes et al. 2010. {``Data Structures for Statistical Computing
in Python.''} In \emph{Proceedings of the 9th Python in Science
Conference}, 445:51--56. Austin, TX.

\leavevmode\vadjust pre{\hypertarget{ref-McNamara:2018}{}}%
McNamara, Amelia, and Nicholas J. Horton. 2018. {``Wrangling Categorical
Data in {R}.''} \emph{The American Statistician} 72 (1): 97--104.
\url{https://doi.org/10.1080/00031305.2017.1356375}.

\leavevmode\vadjust pre{\hypertarget{ref-Meyer:twitter}{}}%
Meyer, Austin. 2021. {``{Austin Meyer on Twitter}.''}
\url{https://twitter.com/austingmeyer/status/1380942918593183744}.

\leavevmode\vadjust pre{\hypertarget{ref-vscode}{}}%
Microsoft. 2023. {``{Visual Studio Code - Code Editing. Redefined}.''}
\url{https://code.visualstudio.com}.

\leavevmode\vadjust pre{\hypertarget{ref-styler}{}}%
Müller, Kirill, and Lorenz Walthert. 2022. \emph{{styler}:
{N}on-Invasive Pretty Printing of r Code}.
\url{https://CRAN.R-project.org/package=styler}.

\leavevmode\vadjust pre{\hypertarget{ref-NASEM:2018}{}}%
National Academies of Science, Engineering, and Medicine. 2018. {``Data
Science for Undergraduates: {O}pportunities and Options.''} National
Academies. \url{https://nas.edu/envisioningds}.

\leavevmode\vadjust pre{\hypertarget{ref-NASEM:2019}{}}%
---------. 2019. {``Reproducibility and Replicability in Science.''}
National Academies. \url{https://nap.edu/25303}.

\leavevmode\vadjust pre{\hypertarget{ref-Nolan:2021}{}}%
Nolan, Deborah, and Sara Stoudt. 2021. \emph{Communicating with Data:
{T}he Art of Writing for Data Science}. Oxford, United Kingdom: Oxford
University Press.

\leavevmode\vadjust pre{\hypertarget{ref-Nolan:2010}{}}%
Nolan, Deboran, and Duncan Temple Lang. 2010. {``Computing in the
Statistics Curriculum.''} \emph{The American Statistician} 64 (2):
97--107. \url{https://doi.org/10.1198/tast.2010.09132}.

\leavevmode\vadjust pre{\hypertarget{ref-Parker:2013}{}}%
Parker, Hilary. 2013. {``{Hilary: the most poisoned baby name in US
history}.''} website.
\url{https://hilaryparker.com/2013/01/30/hilary-the-most-poisoned-baby-name-in-us-history}.

\leavevmode\vadjust pre{\hypertarget{ref-scikit-learn}{}}%
Pedregosa, Fabian, Gaël Varoquaux, Alexandre Gramfort, Vincent Michel,
Bertrand Thirion, Olivier Grisel, Mathieu Blondel, et al. 2011.
{``Scikit-Learn: Machine Learning in Python.''} \emph{Journal of Machine
Learning Research} 12 (Oct): 2825--30.

\leavevmode\vadjust pre{\hypertarget{ref-Postel:1980}{}}%
Postel, Jon. 1980. {``{DoD} Standard Internet Protocol.''} \emph{ACM
SIGCOMM Computer Communication Review} 10 (4): 12--51.
\url{https://datatracker.ietf.org/doc/html/rfc760}.

\leavevmode\vadjust pre{\hypertarget{ref-mosaic:less-volume}{}}%
Pruim, Randall, and Nicholas Horton. 2020. {``Less Volume, More
Creativity -- Getting Started with the Mosaic Package.''}
\url{http://www.mosaic-web.org/mosaic/articles/LessVolume-MoreCreativity.html}.

\leavevmode\vadjust pre{\hypertarget{ref-Riederer:2020}{}}%
Riederer, Emily. 2020. {``Column Names as Contracts.''}
\url{https://emilyriederer.netlify.app/post/column-name-contracts}.

\leavevmode\vadjust pre{\hypertarget{ref-PEP8}{}}%
Rossum, Guido van, Barry Warsaw, and Nick Coghlan. 2023. {``{PEP 8
{\textendash} Style Guide for Python Code}.''} website.
\url{https://peps.python.org/pep-0008}.

\leavevmode\vadjust pre{\hypertarget{ref-rstudio}{}}%
RStudio Team. 2015. \emph{RStudio: Integrated Development Environment
for r}. Boston, MA: RStudio, Inc. \url{http://www.rstudio.com/}.

\leavevmode\vadjust pre{\hypertarget{ref-Saint-Exupery:1984}{}}%
Saint-Exupéry, A. de. 1984. \emph{Airman's Odyssey}. Harcourt Brace
Jovanovich. \url{https://books.google.com/books?id=nIOZdLHReUMC}.

\leavevmode\vadjust pre{\hypertarget{ref-Sandve:2013}{}}%
Sandve, G. K., A. Nektrutenko, J. Taylor, and E. Hovig. 2013. {``Ten
Simple Rules for Reproducible Computational Research.''} \emph{PLoS
Computational Biology}.
\url{https://doi.org/10.1371/journal.pcbi.1003285}.

\leavevmode\vadjust pre{\hypertarget{ref-Schulte:2008}{}}%
Schulte, Carsten. 2008. {``Block Model: An Educational Model of Program
Comprehension as a Tool for a Scholarly Approach to Teaching.''} In
\emph{Proceedings of the Fourth International Workshop on Computing
Education Research}, 149--60. ICER '08. New York, NY, USA: Association
for Computing Machinery. \url{https://doi.org/10.1145/1404520.1404535}.

\leavevmode\vadjust pre{\hypertarget{ref-plotly-r}{}}%
Sievert, Carson. 2020. \emph{Interactive Web-Based Data Visualization
with r, Plotly, and Shiny}. Chapman; Hall/CRC.
\url{https://plotly-r.com}.

\leavevmode\vadjust pre{\hypertarget{ref-Spertus:2021}{}}%
Spertus, Ellen. 2021. {``Best Practices for Writing Code Comments.''}
\url{https://stackoverflow.blog/2021/07/05/best-practices-for-writing-code-comments}.

\leavevmode\vadjust pre{\hypertarget{ref-stackoverflow:reprex}{}}%
Stack Overflow. n.d. {``How to Create a Minimal, Reproducible Example -
Help Center - {S}tack {O}verflow.''}
\url{https://stackoverflow.com/help/minimal-reproducible-example}.

\leavevmode\vadjust pre{\hypertarget{ref-Stegeman:2014}{}}%
Stegeman, Martijn, Erik Barendsen, and Sjaak Smetsers. 2014. {``Towards
an Empirically Validated Model for Assessment of Code Quality.''}
\url{https://dl.acm.org/doi/10.1145/2674683.2674702}.

\leavevmode\vadjust pre{\hypertarget{ref-Stegeman:2016}{}}%
---------. 2016. {``Designing a Rubric for Feedback on Code Quality in
Programming Courses.''}
\url{https://dl.acm.org/doi/10.1145/2999541.2999555}.

\leavevmode\vadjust pre{\hypertarget{ref-Taschuk:2017}{}}%
Taschuk, M., and G. Wilson. 2017. {``Ten Simple Rules for Making
Research Software More Robust.''} \emph{PLOS Computational Biology} 13
(4). \url{https://doi.org/10.1371/journal.pcbi.1005412}.

\leavevmode\vadjust pre{\hypertarget{ref-Theobold:2021}{}}%
Theobold, Allison S., Stacey A. Hancock, and Sara Mannheimer. 2021.
{``Designing Data Science Workshops for Data-Intensive Environmental
Science Research.''} \emph{Journal of Statistics and Data Science
Education} 29 (sup1): S83--94.
\url{https://doi.org/10.1080/10691898.2020.1854636}.

\leavevmode\vadjust pre{\hypertarget{ref-Thomas:2019}{}}%
Thomas, David, and Andrew Hunt. 2019. \emph{The Pragmatic Programmer}.
2nd ed. Boston, MA: Addison Wesley.

\leavevmode\vadjust pre{\hypertarget{ref-Trisovic:2022}{}}%
Trisovic, Ana, Matthew K. Lau, Thomas Pasquier, and Merce Crosas. 2022.
{``A Large-Scale Study on Research Code Quality and Execution.''}
\emph{Science Data}. \url{https://doi.org/10.1038/s41597-022-01143-6}.

\leavevmode\vadjust pre{\hypertarget{ref-unittest}{}}%
{``{unittest --- {Unit} testing framework}.''} 2023. \emph{Python
Documentation}. \url{https://docs.python.org/3/library/unittest.html}.

\leavevmode\vadjust pre{\hypertarget{ref-reticulate}{}}%
Ushey, Kevin, JJ Allaire, and Yuan Tang. 2022. \emph{{reticulate}:
Interface to 'Python'}.
\url{https://CRAN.R-project.org/package=reticulate}.

\leavevmode\vadjust pre{\hypertarget{ref-altair}{}}%
VanderPlas, Jacob, Brian Granger, Jeffrey Heer, Dominik Moritz, Kanit
Wongsuphasawat, Arvind Satyanarayan, Eitan Lees, Ilia Timofeev, Ben
Welsh, and Scott Sievert. 2018. {``Altair: {I}nteractive Statistical
Visualizations for Python.''} \emph{Journal of Open Source Software} 3
(32): 1057.

\leavevmode\vadjust pre{\hypertarget{ref-VanTol:2023}{}}%
VanTol, Alexander. 2023. {``{Python Code Quality: Tools {\&} Best
Practices}.''} website.
\url{https://realpython.com/python-code-quality}.

\leavevmode\vadjust pre{\hypertarget{ref-Vartanian:2022}{}}%
Vartanian, Erica. 2022. {``6 Coding Best Practices for Beginner
Programmers.''}
\url{https://www.educative.io/blog/coding-best-practices\#lowhigh}.

\leavevmode\vadjust pre{\hypertarget{ref-Wickham:2015}{}}%
Wickham, H. 2015. \emph{R Packages}. O'Reilly Media, Incorporated.
\url{https://r-pkgs.org/}.

\leavevmode\vadjust pre{\hypertarget{ref-Wickham:AdvancedR}{}}%
---------. 2019. \emph{Advanced {R}}. 2nd ed. Chapman; Hall/CRC.

\leavevmode\vadjust pre{\hypertarget{ref-testthat}{}}%
Wickham, Hadley. 2011. {``Testthat: {G}et Started with Testing.''}
\emph{The R Journal} 3: 5--10.
\url{https://journal.r-project.org/archive/2011/RJ-2011-002/index.html}.

\leavevmode\vadjust pre{\hypertarget{ref-Wickham:2018}{}}%
---------. 2018. {``Whole Game.''}
\url{https://www.youtube.com/watch?v=go5Au01Jrvs\&t=3s}.

\leavevmode\vadjust pre{\hypertarget{ref-tidy-style}{}}%
---------. 2022. {``{Tidyverse Style Guide}.''}
\url{https://style.tidyverse.org}.

\leavevmode\vadjust pre{\hypertarget{ref-tidyverse}{}}%
Wickham, Hadley, Mara Averick, Jennifer Bryan, Winston Chang, Lucy
D'Agostino McGowan, Romain François, Garrett Grolemund, et al. 2019.
{``Welcome to the {tidyverse}.''} \emph{Journal of Open Source Software}
4 (43): 1686. \url{https://doi.org/10.21105/joss.01686}.

\leavevmode\vadjust pre{\hypertarget{ref-devtools}{}}%
Wickham, Hadley, Jim Hester, Winston Chang, and Jennifer Bryan. 2022.
\emph{{devtools}: {T}ools to Make Developing {R} Packages Easier}.
\url{https://CRAN.R-project.org/package=devtools}.

\leavevmode\vadjust pre{\hypertarget{ref-Wilkinson:2006}{}}%
Wilkinson, L., D. Wills, D. Rope, A. Norton, and R. Dubbs. 2006.
\emph{The Grammar of Graphics}. Statistics and Computing. Springer New
York. \url{https://books.google.com/books?id=NRyGnjeNKJIC}.

\leavevmode\vadjust pre{\hypertarget{ref-Wilson:2017}{}}%
Wilson, Greg, Jennifer Bryan, Karen Cranston, Justin Kitzes, Lex
Nederbragt, and Tracy K. Teal. 2017. {``Good Enough Practices in
Scientific Computing.''} \emph{PLOS Computational Biology} 13 (6):
1--20. \url{https://doi.org/10.1371/journal.pcbi.1005510}.

\leavevmode\vadjust pre{\hypertarget{ref-formatR}{}}%
Xie, Yihui. 2022. \emph{{formatR}: {F}ormat {R} Code Automatically}.
\url{https://CRAN.R-project.org/package=formatR}.

\end{CSLReferences}

\end{document}